\documentclass[11pt, reqno]{article}

\usepackage{jheppub}
\usepackage[nottoc]{tocbibind}

\usepackage[utf8]{inputenc}
\hypersetup{
	colorlinks,
	urlcolor=Maroon,
	linkcolor=Maroon,
	citecolor=Maroon
}

\usepackage[utf8]{inputenc}
\usepackage{natbib}
\usepackage{graphicx}
\usepackage{amsmath}
\usepackage{amssymb}
\usepackage{bbold}
\usepackage{xfrac}

\usepackage{physics}
\usepackage[dvipsnames]{xcolor}
\usepackage{comment}
\usepackage{tabularx}
\usepackage{placeins}
\usepackage{tikz}
\usepackage{subcaption}
\usepackage{enumitem}
\usepackage{empheq}
\usepackage{mathtools}

\newcommand\kett[1]{\ket*{#1} \!\!\: \rangle}
\newcommand\kkett[1]{\ket*{\!\ket*{#1} \!\!\: }}

\newcommand\bbraa[1]{\bra*{\!\!\!\: \bra*{#1}}}
\newcommand{\id}{\mathbb{1}}
\newcommand{\bphi}{\bar{\phi}}

\makeatletter
\newcommand*{\defeq}{\mathrel{\rlap{%
                     \raisebox{0.3ex}{$\m@th\cdot$}}%
                     \raisebox{-0.3ex}{$\m@th\cdot$}}%
                     =}
\makeatother

\title{Conformal Defects from String Field Theory}

\author[a,b,c]{Kasia Budzik,}\emailAdd{kbudzik@perimeterinstitute.ca}
\author[d]{Miroslav Rap\v{c}\'{a}k}\emailAdd{miroslav.rapcak@gmail.com}
\author[a,b,e]{and Jairo M. Rojas}\emailAdd{jmrojash@pucp.edu.pe}

\affiliation[a]{Perimeter Institute for Theoretical Physics, Waterloo, Canada}
\affiliation[b]{Department of Physics $\&$ Astronomy, University of Waterloo, Waterloo, Canada}
\affiliation[c]{CEICO, Institute of Physics of the Czech Academy of Sciences, Prague, Czech Republic}
\affiliation[d]{Center for Theoretical Physics, University of California, Berkeley, USA}
\affiliation[e]{ICTP South American Institute for Fundamental Research, IFT-UNESP, S\~{a}o Paulo, Brazil}

\abstract{Unlike conformal boundary conditions, conformal defects of Virasoro minimal models lack classification. Alternatively to the defect  perturbation theory and the truncated conformal space approach, we employ open string field theory (OSFT) techniques to explore the space of conformal defects. We illustrate the method by an analysis of OSFT around the background associated to the $(1,2)$ topological defect in diagonal unitary minimal models. Numerical analysis of OSFT equations of motion leads to an identification of a nice family of solutions, recovering the picture of infrared fixed points due to Kormos, Runkel and Watts. In particular, we find a continuum of solutions in the Ising model case and 6 solutions for other minimal models. OSFT provides us with numerical estimates of the $g$-function and other coefficients of the boundary state.}

\begin{document}

\maketitle
\addtocontents{toc}{\protect\setcounter{tocdepth}{1}}

\section{Introduction}
\label{ch:Intro}

Rational conformal field theories in two dimensions often admit an exact solution for correlation functions of bulk operators. Examples of such CFTs are diagonal unitary minimal models of the Virasoro algebra $\mathcal{M}_{m+1,m}$ that form a family of models containing the Ising model for $m=3$ and the tricritical Ising model for $m=4$. Using bootstrap techniques, one can solve for structure constants (three-point correlation functions of primary fields) of bulk operators \cite{Dotsenko:1984nm,runkel2000boundary}. With the knowledge of structure constants one can then reconstruct any other correlation function of local operators.

Bulk fields are generally not the only operators in a given theory. One could for example place the theory on a manifold with a boundary (say upper-half plane or unit disk) and impose boundary condition on the bulk fields, i.e. prescribe their behavior near the boundary. On top of the bulk fields, one can then introduce fields living at the boundary. Solving a model thus requires a further knowledge of the boundary spectrum together with boundary structure constants (three-point functions of boundary fields) and bulk-boundary structure constants (two-point functions of one boundary and one bulk field). The solution of unitary diagonal minimal models (simply minimal models from now on) extends to the boundary case. In particular, all conformal boundary conditions have been classified, the spectrum of boundary fields on a given boundary is known \cite{Cardy:1989ir} and the corresponding structure constants were determined in \cite{Runkel:1998he, runkel2000boundary}.

Apart from conformal boundaries, one can introduce a conformal defect that is a one-dimensional object allowing discontinuities or singularities of bulk correlation functions. Conformal defects emerge in various contexts in condensed matter physics \cite{Wong:1994np,Affleck:1995ge,Affleck:1996mm, affleck2008quantum, Saleur:1998hq, Saleur:2000gp, Fendley_2009, Gaiotto:2020fdr}, implement symmetries and order-disorder dualities \cite{Frohlich:2004ef, Frohlich:2006ch, Frohlich:2009gb, Aasen:2020jwb} and constrain bulk and boundary renormalization group flows \cite{Chang:2018iay, Konechny:2019wff, Graham:2003nc}. A natural generalization of conformal defects are conformal interfaces between two different CFTs.

Using the so-called \emph{folding trick} of \cite{Wong:1994np}, a conformal defect in a given CFT can be equivalently described as a conformal boundary in the folded theory $\text{CFT}\otimes\overline{\text{CFT}}$. However, the folded theory is usually not rational with respect to the total Virasoro symmetry and the classification of its conformal boundary conditions, hence conformal defects in the unfolded theory, seems much harder. The only exceptions are defects in the Ising model (whose folded theory is equivalent to the free boson on $S^1/\mathbb{Z}_2$) \cite{Oshikawa:1996ww, Oshikawa:1996dj}, defects in the Lee-Yang model and interfaces between these two \cite{Quella:2006de}. Even though general conformal defects are far from being fully understood, there are two families of defects, called \emph{topological} (totally transmissive) and \emph{factorizing} (totally reflective), that allow classification. Moreover, in some specific cases, the folded theory might turn out to be rational with respect to an extended symmetry and a subset of boundary conditions preserving this extended symmetry can be constructed \cite{Gang:2008sz, Makabe:2017ygy, Fredenhagen:2006qw}.

In addition to bootstrap methods, one can study the space of defects in a given theory by deformations and the renormalization group flow. This approach was implemented by Kormos, Runkel and Watts in \cite{Kormos:2009sk} who studied a two-parameter family of deformations of a particular topological defect in minimal models $\mathcal{M}_{m+1,m}$ for general $m$. Using the conformal perturbation theory and the truncated conformal space approach together with various exact insights, they identified a continuum of IR fixed points in the Ising model $m=3$, and six IR fixed points for other values of $m$. The structure of the fixed points is reproduced in figure \ref{fig:all-flows}. Five of the fixed points can be identified with topological and factorizing defects, while the last one was argued to be a new non-trivial conformal defect which is not a superposition of factorizing and topological defects. Very little is known about this mysterious defect besides the leading large $m$ behaviors of the $g$-function \cite{Kormos:2009sk} and its reflection coefficient \cite{Makabe:2017uch}.

As discussed in this note, there exists yet another method that has the potential to identify and study new conformal defects. This method is based on open string field theory (OSFT) \cite{Witten:1985cc} originally developed to address non-perturbative aspects of string theory and provide its background-independent formulation. Starting with a given OSFT background (a configuration of branes on which open strings end), one can write down an action for all the open-string modes and look for solutions of the corresponding classical equations of motion. Such solutions can be then interpreted as new OSFT backgrounds \cite{Sen:1999mh,Sen:1999xm}. From the worldsheet perspective, an OSFT background corresponds to a choice of a conformal boundary condition of the bulk CFT. Solutions of the OSFT equations of motion can be thus interpreted as new conformal boundary conditions and one could hope that OSFT might be used to study new conformal boundaries and defects.

The space of open-string states in a given background can be roughly identified with the boundary spectrum of the corresponding boundary. A \textit{string field} is then a linear combination of open-string states and the OSFT action is a cubic polynomial in components of the string field.  Writing the full action and finding its saddle points is an enormously hard problem \cite{Schnabl:2005gv,Okawa:2006vm,Erler:2009uj,Erler:2014eqa,Erler:2019fye,Erler:2019vhl}. Luckily, a level truncation method was developed to look for such solutions numerically \cite{Rastelli:2000iu,Sen:1999nx,Taylor:2000ek,Moeller:2000xv,Moeller:2000jy,Sen:1999xm,Kudrna:2016ack,Kudrna:2014rya,Kudrna:2018mxa,Kudrna:2019xnw}. Level truncation is based on a restriction of the string field to components with conformal weight smaller than some level $L$ and with all the other components set to zero. One can then look for truncated classical solutions which are stable under the level increase. Obviously, one can expect only a small subset of solutions to be visible by such a method but luckily enough one can indeed find non-trivial solutions.

By coupling a CFT whose boundary conditions we want to study to the world-sheet of OSFT, one can use OSFT to search for new conformal boundary conditions. This method has already been implemented in various contexts in \cite{Kudrna:2014rya, Schnabl:2019oom, Vosmera2}. Here, we are going to apply it to a less explored problem of conformal defects \footnote{A different connection between topological defects and OSFT was also investigated in \cite{Kojita:2016jwe}, where topological defects were shown to relate OSFT solutions for different BCFTs.} which is possible due to the folding trick. Motivated by the above-mentioned work of Kormos, Runkel and Watts, we study an OSFT analogue of their setup. Specifically, we consider the topological defect $D_{(1,2)}$ as our starting background. The defect spectrum contains in particular two relevant primary fields labeled as $\phi_{(1,3)(1,1)}$ and $\phi_{(1,1)(1,3)}$ that were used in \cite{Kormos:2009sk} to deform the action and search for RG fixed points. We restrict to a subset of fields closed under the operator product expansion (OPE) and generated by fields $\phi_{(1,3)(1,1)}$ and $\phi_{(1,1)(1,3)}$ together with their descendants. One can then use the results of \cite{Makabe:2017uch} for the defect-defect structure constants to write down the OSFT action up to level 4 and search numerically for its saddle points.

Already at level one, i.e. considering the string field with only three components corresponding to the highest weight states $\ket{0}$, $\ket*{\phi_{(1,3)(1,1)}}$ and $\ket*{\phi_{(1,1)(1,3)}}$, one recovers the structure of fixed points from \cite{Kormos:2009sk} obtained by a combination of conformal perturbation theory and the truncated conformal space approach. In particular, one recovers a continuum of solutions for the Ising model and 6 discrete solutions for other unitary minimal models. Improving the solutions to higher levels and extrapolating in $L$, one can give a numerical prediction for the $g$-functions. We find a nice agreement with the $g$-function of defects identified in \cite{Kormos:2009sk} and give a prediction for the $g$-function of the new conformal defect $C$. The continuum of solutions in the Ising model have equal value of the $g$-function but different defects can be distinguished by other coefficients of the boundary state. These can be calculated from the OSFT solution using generalized Ellwood invariants \cite{Kudrna:2012re}.

This paper is organized as follows: Section \ref{ch:MM} reviews some basics of conformal boundaries and defects in unitary minimal models together with the main results of Kormos, Runkel and Watts \cite{Kormos:2009sk}. In section \ref{ch:OSFT}, we state a prescription for the level truncation method that will hopefully be accessible also to people outside of the OSFT community. Finally, section \ref{ch:Results} illustrates the method on the OSFT analogue of the Kormos-Runkel-Watts setup and gives a numerical prediction for the $g$-functions of the involved conformal defects, including the defect $C$. Our analysis only scratches the surface of possible use of OSFT. We mention a few possible directions in the conclusion.

\section{Minimal models}
\label{ch:MM}

Unitary minimal models of the Virasoro algebra with diagonal partition function (\emph{minimal models} from now on) form a special class of two-dimensional conformal field theories $\mathcal{M}_{m+1,m}$ parametrized by an integral parameter $m\geq 3$. Some minimal models can be identified with a continuum limit of a lattice model at a critical point, such as the Ising model for $m=3$ and the tricritical Ising model for $m=4$.

Hilbert spaces of minimal models are built out of a finite number of irreducible highest-weight representations of the Virasoro algebra (referred to as conformal families). The finite set of conformal families $R_{\alpha}$ in $\mathcal{M}_{m+1,m}$ is parametrized by pairs of integers called Kac labels $\alpha\in \mathcal{I}_m$:
\begin{equation}
    \mathcal{I}_{m} = \qty{ (r,s)\in\mathbb{Z}^2 \; | \; 1\leq r\leq m-1, \; 1\leq s\leq m }/\sim,
    \label{eq:Kac-labels}
\end{equation}
where we identify $(r,s)\sim (m\!-\!r, m\!+\!1\!-\!s)$. The central charge of the theory and conformal weights of primary operators are
\begin{align}
c &= 1 - \frac{6}{m(m+1)}, \\
h_{r,s} &= \frac{((m+1)r-ms)^2-1}{4m(m+1)}, \quad (r,s)\in\mathcal{I}_m.
\end{align}
The Hilbert space is then given in terms of a direct sum
\begin{equation}
    \bigoplus_{\alpha\in\mathcal{I}_m} R_{\alpha} \otimes \bar{R}_{\alpha},
\end{equation}
where $R_{\alpha}$ and $\bar{R}_{\alpha}$ are conformal families associated to the holomorphic and antiholomorphic components of the stress-energy tensor. 

\subsection{Conformal boundaries}


Next, let us place our CFT on a manifold with a boundary, for example the upper-half plane (UHP). A CFT together with a choice of a consistent conformal boundary condition is referred to as boundary conformal field theory (BCFT). A boundary condition in a BCFT is represented by a boundary state $\kkett{B}$ encoding the behavior of bulk fields near the boundary, i.e.
\begin{equation}
\bra*{\!\! \bra*{B} \!\!\: } \ket{\phi}= \expval{\,\phi(0,0)\,}^{\text{disk}}_B = 4^{h} \expval{\, \phi(i,-i) \,}^{\text{UHP}}_B,
\label{bstate}
\end{equation}
for any spinless bulk primary operator $\phi(z,\bar{z})$ of conformal weights $(h,h)$. From each boundary state $\kkett{B}$, one can extract a useful quantity called the \textit{ground state degeneracy} or the \textit{$g$-function} \cite{Affleck:1991tk} corresponding to the coefficient in front of the vacuum state $\ket{0}$, i.e.
\begin{equation}
g(B) = \bra*{\!\! \bra*{B} \!\!\: } \ket{0} = \expval{ \id }_B.
\label{eq:g(B)}
\end{equation}
The $g$-function is a BCFT analog of the $c$-function and decreases (or remains constant) along the boundary RG flow \cite{Affleck:1992ng, Friedan:2003yc}.

Introducing a boundary obviously breaks half of the conformal symmetry that displaces the boundary. A boundary condition is called \textit{conformal} if it preserves the remaining conformal transformations. The condition of conformal invariance is equivalent to the vanishing of the diagonal component of the stress-energy tensor at the boundary
\begin{equation}
T(z) - \bar{T}(\bar{z})=0, \quad z\in\mathbb{R}.
\label{eq:T-T}
\end{equation}
The constraint translates to the condition on the boundary state
\begin{equation}
(L_n-\bar{L}_{-n}) \kkett{B} = 0, \quad \forall n\in\mathbb{Z}.
\label{eq:L-L}
\end{equation}
The space of solutions of the above equation is spanned by the so-called Ishibashi states \cite{Ishibashi:1988kg}, which are in a one-to-one correspondence with diagonal bulk primaries $\phi_{\alpha}$. If we denote the orthonormal basis of an irreducible representation of the Virasoro algebra $R_{\alpha}$ by  $\qty{\ket{\phi_{\alpha}, n}}_n$, the corresponding Ishibashi state is explicitly given by
\begin{equation}
\kett{\phi_{\alpha}} = \sum_{n} \ket{\phi_{\alpha}, n}\otimes \overline{\ket{\phi_{\alpha}, n}} , \qquad \alpha\in\mathcal{I}_m .
\label{eq:Ishibashi}
\end{equation}
Boundary state $\kkett{B}$ is then given in terms of a linear combination of Ishibashi states with coefficients encoded in overlaps (\ref{bstate}).


As mentioned above, any linear combination of Ishibashi states satisfies the condition of conformal invariance (\ref{eq:L-L}). However, there are other consistency conditions, which further restrict the allowed conformal boundary states. In particular, we have the Cardy condition arising from the consistency of the theory when placed on the cylinder \cite{Cardy:1989ir, Cardy:2004hm}. States which satisfy this condition can be built up from a set of elementary states called \textit{Cardy states}. In case of minimal models, Cardy states are again in a one-to-one correspondence with bulk primaries and are explicitly given by
\begin{equation}
\kkett{\phi_{\alpha}} = \sum_{\beta\in\mathcal{I}_m} \, \frac{S_{\alpha\beta}}{\sqrt{S_{0\beta}}} \, \kett{\phi_{\beta}} , \qquad \alpha\in\mathcal{I}_m ,
\label{eq:Cardy-state}
\end{equation}
where zero denotes the $(1,1)$ Kac label and $S_{\alpha\beta}$ are the entries of the modular $S$-matrix of the minimal model $\mathcal{M}_{m+1,m}$:
\begin{equation}
S_{(r_1,s_1)(r_2,s_2)} = \sqrt{\frac{8}{m(m+1)}} \, (-1)^{1+r_1s_2+r_2s_1} \, \sin\qty(\frac{\pi\, (m+1)\, r_1\, r_2}{m}) \sin\qty(\frac{\pi\, m\, s_1\, s_2}{m+1}).
\label{eq:Smatrix}
\end{equation}
The most general conformal boundary condition is then a superposition of elementary Cardy states
\begin{equation}
\kkett{B} = \sum_{\alpha\in\mathcal{I}_m} n_\alpha \kkett{\phi_{\alpha}},
\end{equation}
with $n_\alpha$ being non-negative integers. 

BCFT also contains local fields living at the boundary. The boundary spectrum associated to the boundary state $\kkett{\phi_{\alpha}}$ decomposes into irreducible representations $R_{\gamma}$ of a single Virasoro algebra preserved by the boundary. The multiplicity of a given conformal family $R_{\gamma}$ is given by the integer $\mathcal{N}_{\alpha\alpha}^{\gamma}$ appearing in the fusion rules of bulk operators \cite{Cardy:1989ir}
\begin{equation}
    \phi_{\alpha} \times \phi_{\beta} = \sum_{\gamma} \mathcal{N}_{\alpha\beta}^{\gamma} \; \phi_{\gamma}.
    \label{eq:fusion-rules}
\end{equation}
In case of unitary diagonal minimal models, the fusion rules read
\begin{equation}
   \phi_{(r_1,s_1)} \times \phi_{(r_2,s_2)} = \sum_{\substack{k=1+|r_1-r_2| \\ k+r_1+r_2 \text{ odd}}}^{k_{max}} \;\; \sum_{\substack{l=1+|s_1-s_2| \\ l+s_1+s_2 \text{ odd}}}^{l_{max}} \; \phi_{(k,l)},
    \label{eq:MM-fusion-rules}
\end{equation}
where $k_{max}=\min \qty(r_1\!+\!r_2\!-\!1, 2m\!-\!1\!-\!r_1\!-\!r_2)$, $l_{max}=\min \qty(s_1\!+\!s_2\!-\!1, 2m\!-\!1\!-\!s_1\!-\!s_2)$. The coefficients $\mathcal{N}_{\alpha\beta}^{\gamma}$ equal 0 or 1 depending on whether or not the conformal family of $\phi_{\gamma}$ appears on the right hand side of (\ref{eq:MM-fusion-rules}).



\subsection{Conformal defects}
\label{subch:defects}

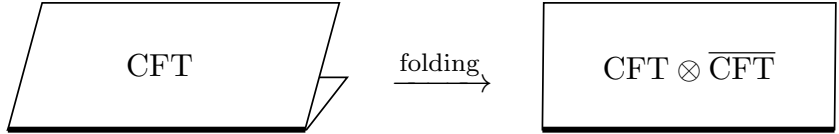
\begin{figure}[t]
\centering

\tikzset{every picture/.style={line width=0.65pt}} 

\begin{tikzpicture}[x=0.75pt,y=0.75pt,yscale=-0.8,xscale=0.8]

\draw   (102.99,262.4) -- (291.3,262.4) -- (265.61,295.4) -- (77.3,295.4) -- cycle ;
\draw  [fill={rgb, 255:red, 255; green, 255; blue, 255 }  ,fill opacity=1 ] (98.94,215.4) -- (286.3,215.4) -- (264.66,295.4) -- (77.3,295.4) -- cycle ;
\draw [line width=2.25]    (77.3,295.4) -- (264.89,295.4) ;

\draw  [fill={rgb, 255:red, 255; green, 255; blue, 255 }  ,fill opacity=1 ] (415.14,215.4) -- (599.3,215.4) -- (598.46,295.4) -- (414.3,295.4) -- cycle ;
\draw [line width=2.25]    (414.3,295.4) -- (598.46,295.4) ;

\draw (351,258.4) node [scale=1.2]  {${\displaystyle \xrightarrow{{\mathrm{folding}}}}$};
\draw (173,254.4) node [scale=1.1]  {$\mathrm{CFT}$};
\draw (506.8,255.4) node [scale=1.1]  {$\mathrm{CFT} \otimes \mathrm{\overline{CFT}}$};
\end{tikzpicture}
\bigskip
\caption{ Folding trick relates a defect CFT to a boundary in a product CFT.}
\label{fig:folding}
\end{figure}

A defect in a two-dimensional theory is a one-dimensional line along which correlation functions of bulk operators might be singular or discontinuous. By the folding trick \cite{Wong:1994np}, defects in a given CFT are equivalent to boundary conditions in a folded theory $\text{CFT}\otimes\overline{\text{CFT}}$, where $\overline{\text{CFT}}$ denotes a CFT with holomorphic and antiholomorphic dependencies switched. We can also define the $g$-function of a defect as the $g$-function of the corresponding boundary condition. The folding trick is illustrated in figure \ref{fig:folding}.


Analogously to conformal boundary conditions, a defect is called \textit{conformal} if it preserves a subset of conformal transformations which do not displace it. The condition of conformal invariance is equivalent to the continuity of the diagonal component of the stress energy tensor, i.e. $T(z) - \bar{T}(\bar{z})$, across the defect. The difficulty of understanding general conformal defects is related to the fact that product CFTs are generally much harder to study. For example, even if the spectrum of defect primaries is finite in the original theory, it may contain infinitely many primaries with respect to the Virasoro algebra generated by modes of the total stress-energy tensor of the folded system. There exist two families of conformal defects, \emph{topological} and \emph{factorizing}, which satisfy stronger conditions and as a result admit classification at least in the case of minimal models.

\subsubsection{Factorizing defects}


First, let us consider factorizing defects satisfying
\begin{equation}
T(z)-\bar{T}(\bar{z})=0, \quad z\in\mathbb{R}.
\end{equation}
A simple example of a factorizing defect is a system of two independent theories on opposite sides of the defect with a choice of a boundary condition for each of them. More generally, factorizing defects are equivalent to a superposition of products of conformal boundaries
\begin{equation}
F = \sum_{ \alpha, \beta\in\mathcal{I}_m} n_{\alpha\beta} \kkett{\phi_\alpha} \! \bbraa{\phi_\beta},
\label{eq:facto-defect2}
\end{equation}
where the coefficients $n_{\alpha\beta}$ are non-negative integers. Correlation functions in a system with a factorizing defect simply factorize into the sum of products of contributions from both half-planes. In particular, the $g$-function of the above defect is given by
\begin{equation}
g(F) = \sum_{ \alpha, \beta\in\mathcal{I}_m } n_{\alpha\beta}\frac{S_{\alpha 0} \, S_{\beta 0}}{S_{00}} .
\label{eq:g(F)}
\end{equation}

\subsubsection{Topological defects}

Topological defects form another class of conformal defects such that both $T(z)$ and $\bar{T}(\bar{z})$ are continuous across the defect. As a consequence, they can be continuously deformed (away from insertions of other operators) without changing correlation functions.

Topological defects in minimal models are labeled by Kac labels $\alpha\in\mathcal{I}_m$ \cite{Petkova:2000ip, Petkova:2013yoa}. In terms of operators acting on bulk fields, they can be written in the form
\begin{equation}
    D_{\alpha} = \sum_{\beta\in\mathcal{I}_m} \frac{S_{\alpha \beta}}{S_{0 \beta }} \, \text{id}_{R_\beta\otimes\bar{R}_\beta},
\end{equation}
where $\text{id}_{R_\beta\otimes\bar{R}_\beta}$ projects onto $R_\beta\otimes\bar{R}_\beta$. 
The $g$-function of the above defect equals
\begin{equation}
g(D_{\alpha}) = \bra{0} D_{\alpha}\ket{0}= \frac{S_{\alpha 0}}{S_{00}}, \quad \alpha\in\mathcal{I}_m.
\label{eq:g(D)}
\end{equation}
Defect fields are labelled by a pair of Kac labels $\phi_{(r,s)(k,l)} \equiv \phi_{(r,s)}\otimes\bar{\phi}_{(k,l)}$ indicating representations of the Virasoro algebras associated to $T(z)$ and $\bar{T}(\bar{z})$. The spectrum of defect fields on a topological line $D_\alpha$ is given by \cite{Kormos:2009sk}
\begin{equation}
\bigoplus_{\beta,\gamma\in\mathcal{I}_m} \qty[\, R_{\beta} \otimes \bar{R}_{\gamma} \,]^{ \oplus \, \mathcal{M}_{\beta\gamma}},
\label{eq:defect-spectrum}
\end{equation}
where $\mathcal{M}_{\beta\gamma}$ is the multiplicity of the representation $R_{\beta} \otimes \bar{R}_{\gamma}$ given by
\begin{equation}
 \mathcal{M}_{\beta\gamma}= \sum_{\delta\in\mathcal{I}_m} \mathcal{N}_{\beta\gamma}^{\delta} \, \mathcal{N}_{\alpha\alpha}^{\delta}
\label{eq:Ms}
\end{equation}
and $\mathcal{N}_{\alpha\beta}^{\gamma}$ are again the integers appearing in the fusion rules (\ref{eq:fusion-rules}) of the bulk theory.

\subsection{KRW setup}


In this section, we summarize the results of Kormos, Runkel and Watts \cite{Kormos:2009sk}. In the rest of the paper we are going to recover the same picture from a different perspective based on string field theory.

The starting point of the setup considered in \cite{Kormos:2009sk} is the topological defect $D_{(1,2)}$ with the $g$-function
\begin{equation}
    g_0 \equiv g\qty(D_{(1,2)}) = -2 \cos\qty(\frac{\pi m}{m+1})
\end{equation}
placed in a minimal model $\mathcal{M}_{m+1,m}$ for general value of $m$. The spectrum of defect fields is given by (\ref{eq:defect-spectrum}) and contains in particular two relevant fields $\phi_{(1,3)(1,1)}$ and $\phi_{(1,1)(1,3)}$ of conformal dimension $h=(m\!-\!1)/(m\!+\!1)$.  One can thus study the two-parameter space of deformations of $D_{(1,2)}$ by such fields
\begin{equation}
\lambda_l \, \phi_{(1,3)(1,1)} + \lambda_r \, \phi_{(1,1)(1,3)}.
\end{equation}

\begin{figure}[t]
\centering
\begin{subfigure}[t]{0.48\textwidth}
    \centering
	\includegraphics[width=\textwidth]{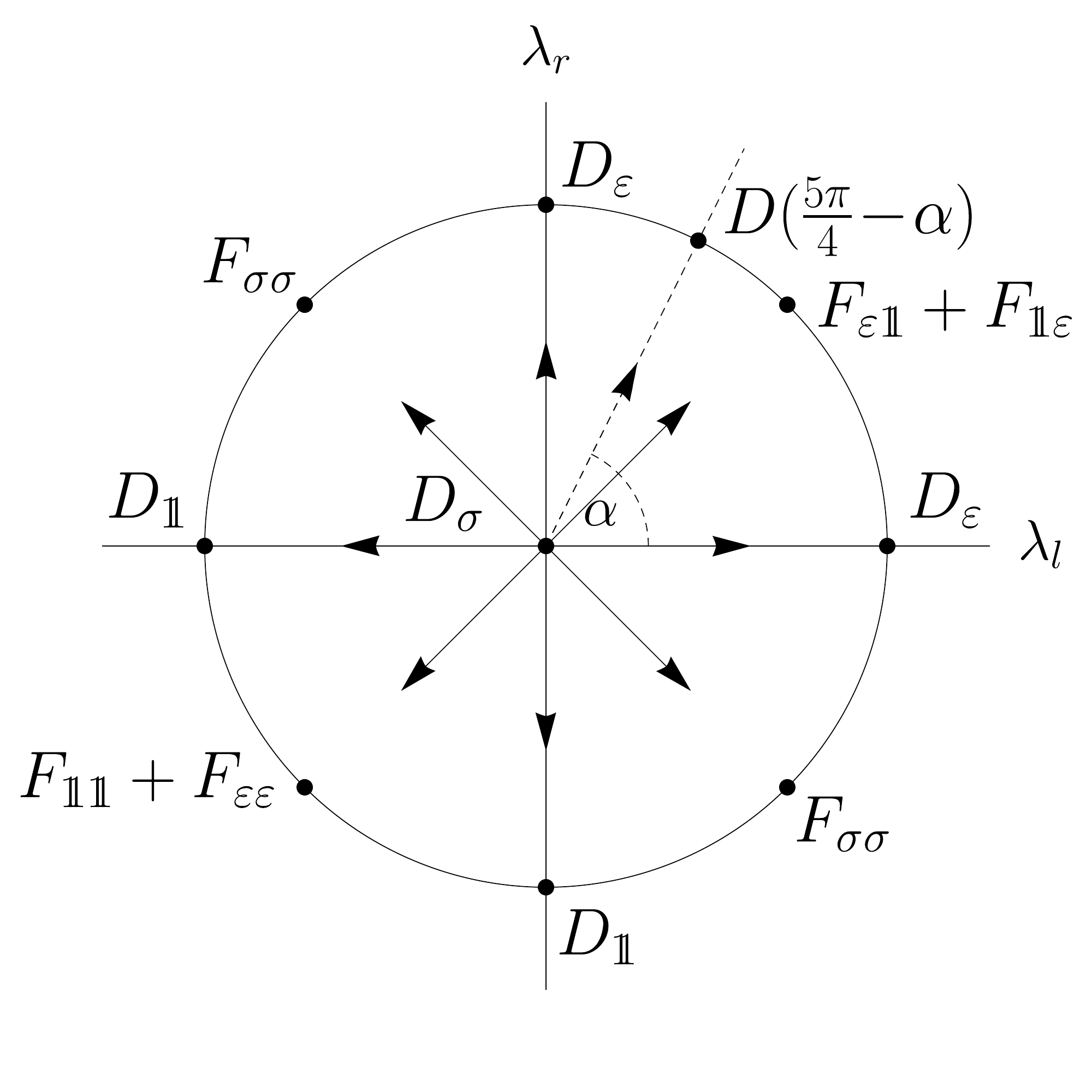}
	\caption{In the Ising model case, there is a continuum of fixed points with particular points corresponding to topological and factorizing defects.}
	\label{fig:ising-flows}
\end{subfigure}
\hspace{0.1cm}
\begin{subfigure}[t]{0.48\textwidth}
    \centering
	\includegraphics[width=\textwidth]{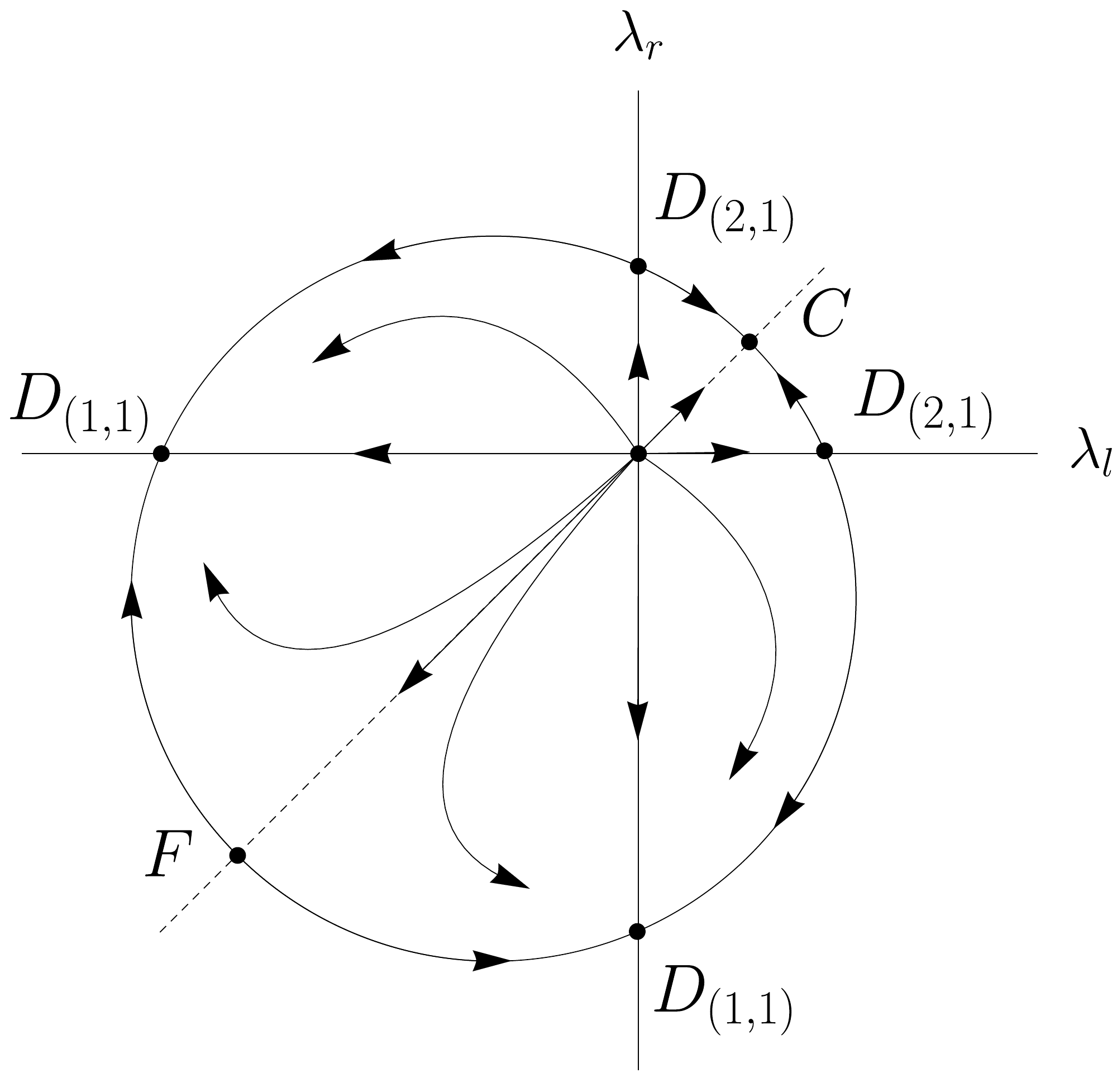}
    \caption{In the $m>3$ minimal model case, there are 6 fixed points. }
    \label{fig:largem-flows}
\end{subfigure}
\caption{Structure of RG fixed points reproduced from \cite{Kormos:2009sk}.}
\label{fig:all-flows}
\end{figure}

Based on conformal perturbation theory, truncated conformal space approach and exact insights, KRW proposed the space of RG flows depicted in figure \ref{fig:all-flows}. The picture differs qualitatively in the case of $m>3$ minimal models and the Ising model with $m=3$.

\subsubsection{$m>3$ minimal models}

The space of defect deformations for $m>3$ minimal models contains six IR fixed points associated to new conformal defects. Three of the fixed points are already visible at the leading order in conformal perturbation theory
\begin{alignat}{3}
    (i) \quad & \lambda_l=\lambda^*, && \quad \lambda_r=0 \\
    (ii) \quad & \lambda_l=0, && \quad \lambda_r=\lambda^* \\
    (iii) \quad & \lambda_l=\lambda^*, && \quad \lambda_r=\lambda^*,
\end{alignat}
where $\lambda^* = (1-h)/C^{\phi}_{\phi\phi}$ and $C^{\phi}_{\phi\phi}=4/3 + \mathcal{O}(1/m)$\footnote{ $C^{\phi}_{\phi\phi}$ is the defect structure constant of three defect fields $\phi\equiv\phi_{(1,3)(1,1)}$, also equal to the defect structure constant $C^{\bar{\phi}}_{\bar{\phi}\bar{\phi}}$ of three defect fields $\bar{\phi}\equiv\phi_{(1,1)(1,3)}$. Our normalization differs from \cite{Kormos:2009sk}, where the two point function is $d_{\phi\phi}=1$ and $C^{\phi}_{\phi\phi}=\sqrt{8/3}+\mathcal{O}(1/m)$. Relevant defect structure constants are listed in appendix \ref{app:StructureConstants}. }. The fixed points $(i)$ and $(ii)$ were identified with the topological defect $D_{(2,1)}$. The fixed point $(iii)$ was conjectured to be a new nontrivial defect, denoted as $C$, which is not a superposition of topological and factorizing defects. The value of the $g$-function in the leading-order conformal perturbation theory reads
\begin{equation}
    \log g(\lambda_l,\lambda_r) - \log g_0  = -\pi^2 d_{\phi\phi} \, (\lambda_l^2+\lambda_r^2) y + \frac{2\pi^2}{3} C^{\phi}_{\phi\phi} d_{\phi\phi} \, (\lambda_l^3+\lambda_r^3) + \mathcal{O}(\lambda^4),
\label{eq:gpert}
\end{equation}
where $y \equiv 1-h=2/(m+1)$. In particular, for the $C$ defect in the large $m$ limit, the $g$-function equals
\begin{align}
\log g(C)-\log g_0 =& -\frac{\pi^2}{4} y^3 + \mathcal{O}(y^4) .
\label{eq:g(C)}
\end{align}

The TCSA analysis reveals in total six fixed points including the above three. The remaining fixed points were identified with the $D_{(1,1)}$ defect (the trivial line) and a factorizing defect conjectured to be the superposition
\begin{equation}
F = \sum_{r=1}^{m-1} \kkett{\phi_{(r,1)}} \! \bbraa{\phi_{(r,1)}}
\label{eq:defectF}
\end{equation}
with the $g$-function equal to
\begin{equation}
g(F) = \sum_{r=1}^{m-1} \frac{\qty(S_{(r,1)0})^2}{S_{00}}.
\end{equation}

\subsubsection{The Ising model}
\label{subsubsec:KRWIsing}

The case of the Ising model qualitatively differs from the $m>3$ minimal model case as the space of defect deformations contains a continuum of fixed points. In order to describe the endpoints of the RG flows, let us first recall the general classification of conformal defects in the Ising model.

The Ising minimal model $\mathcal{M}_{4,3}$ has three distinct Kac labels labelling bulk fields $\mathbb{1}\equiv \phi_{(1,1)}$, $\sigma\equiv \phi_{(1,2)}$ and $\varepsilon\equiv \phi_{(1,3)}$. Therefore, there are three topological defects $D_\mathbb{1}$, $D_\sigma$, $D_\varepsilon$ and nine elementary factorizing defects corresponding to different products of boundary states $\kkett{\mathbb{1}}, \kkett{\sigma}, \kkett{\varepsilon}$. In addition, all the other conformal defects in the Ising model are known as well \cite{Oshikawa:1996ww, Oshikawa:1996dj}. The classification can be carried out in terms of conformal boundary conditions of the folded theory which can be identified with the free boson propagating on the orbifold $S^1/\mathbb{Z}$ of radius 1 in the normalization $\phi (z)\phi (w)\sim -\frac{1}{4}\log (z-w)$. The model obviously admits two families of conformal boundary conditions parameterized by continuous parameters: Dirichlet boundary conditions $\kkett{D(\varphi)}$, $\varphi\in (0,\pi)$ fixing the value of the free boson to be at a given point of $S^1/\mathbb{Z}$, and Neumann boundary conditions $\kkett{N(\tilde{\varphi})}$, $\tilde{\varphi}\in \qty(0,\pi/2)$, differing by the value of the background $U(1)$ gauge field on $S^1/\mathbb{Z}$. At the endpoints of the intervals, the boundary states split into two elementary states (fractional branes) $\kkett{D(0)_\pm}, \kkett{D(\pi)_\pm}$ and $\kkett{N(0)_\pm}, \kkett{N(\pi)_\pm}$ that can be identified with various elementary factorizing defects.

The perturbative analysis of defect deformations is qualitatively different due to the vanishing of all the three-point functions involving fields $\phi=\phi_{(1,3)(1,1)}$ and $\bar{\phi}=\phi_{(1,1)(1,3)}$. The only nonzero defect structure constant in the Ising model is $C^{\Phi}_{\phi\bar{\phi}}$, where $\Phi\equiv\phi_{(1,3)(1,3)}$. However, the combination appearing in the leading-order perturbation calculation vanishes: $C^{\Phi}_{\phi\bar{\phi}}+C^{\Phi}_{\bar{\phi}\phi}=0$. At next-to-leading order conformal perturbation theory, one finds a contribution proportional to $-d_{\phi \phi}^2(\lambda_r^2+\lambda_l^2)\lambda_r$ in the $\beta$ function of $\lambda_r$ and $-d_{\phi \phi}^2(\lambda_r^2+\lambda_l^2)\lambda_l$ in the $\beta$ function of $\lambda_l$. These terms come from the only non-vanishing four-point functions $\langle \phi \phi \phi \phi\rangle$, $\langle \bar{\phi}\bar{\phi} \bar{\phi}\bar{\phi}\rangle$ and $\langle \phi \phi \bar{\phi}\bar{\phi}\rangle$, all proportional to $d_{\phi\phi}^2$. We can see that this next-to-leading-order conformal perturbation theory produces a rotationally invariant result with a continuum of fixed points. On top of that, the existence of the exactly marginal defect field $\Phi$, which can be used to trigger marginal deformation of the IR fixed points, supports the expectation of a continuum of IR fixed points (see \cite{Kormos:2009sk} and \cite{Fendley_2009} for a detailed discussion).  

Let us parametrize the fixed points by the angle $\alpha$ defined by
\begin{align}
    \tan\alpha = \frac{\lambda_r}{\lambda_l},
\end{align}
with $\alpha\in\qty(-\frac{\pi}{2},\frac{\pi}{2})$ for $\lambda_l>0$ and $\alpha\in\qty(\frac{\pi}{2},\frac{3\pi}{2})$ for $\lambda_l<0$. The flows, illustrated in figure \ref{fig:ising-flows}, then interpolate between defects \cite{Kormos:2009sk,Fendley_2009}
\begin{align}
    D_{\sigma}=D_{(1,2)} \longrightarrow D(5\pi/4-\alpha),
\end{align}
where $D(5\pi/4-\alpha)$ is the defect associated to the Dirichlet boundary condition $\kkett{D(5\pi/4-\alpha)}$ in the folded picture. Certain points in the continuum can be identified with topological or factorizing defects
\begin{eqnarray}
D(0) &=& \kkett{\id}\!\!\bbraa{\id} + \kkett{\varepsilon}\!\!\bbraa{\varepsilon} \nonumber \\
D(\pi/4) &=& D_{\id} \nonumber \\
D(\pi/2) &=& \kkett{\sigma}\!\!\bbraa{\sigma} \label{eq:reff} \\
D(3\pi/4) &=& D_{\varepsilon} \nonumber \\
D(\pi) &=& \kkett{\id}\!\!\bbraa{\varepsilon} + \kkett{\varepsilon}\!\!\bbraa{\id} \nonumber.
\end{eqnarray}
Note that the the formula (\ref{eq:defectF}) for the factorizing defect $F$ for $m=3$ holds also in the Ising model and the defect $C$ becomes simply the combination $D(\pi)$.

All conformal defects of the $D(5\pi/4-\alpha)$ family have equal $g$-functions $g=1$. They however differ by correlation functions of bulk fields in the presence of the defect. They can be distinguished for example by the following correlation function
\begin{equation}
\frac{1}{2} \qty(\bra{\varepsilon} D(5\pi/4-\alpha)\ket{\id} + \bra{\id} D(5\pi/4-\alpha)\ket{\varepsilon} )= \cos (2(5\pi/4-\alpha)) = \sin (2\alpha).
\label{eq:sin2x}
\end{equation}
This formula easily follows from the identification of the orbifold-boson vertex operator $\cos 2\phi$ with the combination $\frac{1}{2}\left (\epsilon \otimes \id+\id\otimes \epsilon \right )$ of the two copies of the Ising model and the fact that the value of $\phi$ is fixed to $5\pi/4-\alpha$ for the Dirichlet boundary condition $\kkett{D(5\pi/4-\alpha)}$. It is easy to check that this value agrees with the one obtained by performing the overlaps of (\ref{eq:sin2x}) for special values of $\alpha$ from (\ref{eq:reff}).

\section{Open string field theory}
\label{ch:OSFT}

\subsection{Generalities}

In this section, we very briefly review the basics of classical open string field theory (OSFT). For a proper introduction, we refer for example to \cite{maccaferri2006basics, Rastelli:2005mz, Okawa:2012ica, Kudrna:2018mxa, Erler:2019vhl} and the references therein. The new developments in the field of string field theory include among many \cite{Erbin:2019spp, Masuda:2019rgv, Masuda:2020tfa, Erler:2019xof, Sen:2020oqr, Koyama:2020qfb, Okawa:2020llq, Erbin:2020eyc, Kojita:2016jwe}.


OSFT was introduced as a proposal for a non-perturbative formulation of bosonic string theory in \cite{Witten:1985cc}. String theory is described from the worlsheet perspective in terms of a matter CFT of central charge $c^{\text{matter}}=26$ tensored with the $bc$-ghost system of central charge $c^{\text{gh}}=-26$ such that the total central charge vanishes. We refer to the choice of the matter CFT as the choice of the closed-string background since closed-string excitations are associated to bulk primaries.

The worldsheet of an open string is on the other hand described in terms of a BCFT. The choice of the boundary condition for the combined matter-ghost CFT (encoded by a boundary state $\kkett{B_{0}}$) is referred to as a choice of the open-string background with the spectrum of open strings associated to the boundary spectrum of the corresponding BCFT.

The string field $\Psi$ is defined as a linear combination of fields appearing in the boundary spectrum of the system. In classical OSFT, we restrict to string fields of ghost number\footnote{The $bc$-ghost system admits a $U(1)$ symmetry with respect to which $b$ has charge (ghost number) $-1$ and $c$ has charge $+1$.} equal $+1$. Concrete examples of string fields will be discussed in later sections. The classical action of OSFT is a functional on the space of string fields. Critical points of the OSFT action are expected to correspond to new OSFT backgrounds associated to new boundary conditions of the world-sheet CFT \cite{Sen:1999mh, Sen:1999xm}.


Let us now review the construction of the OSFT action. The action has a Chern-Simons-like form\footnote{Throughout the text, we omit the coupling-constant dependence that would only complicate the expressions bellow. Note in particular that equations of motion are independent of the coupling constant.}
\begin{equation}
S[\Psi] =  - \qty (\frac{1}{2}  \bra{\Psi} \ket{Q_B \Psi}  + \frac{1}{3}  \bra{\Psi} \ket{\Psi \ast \Psi} ) ,
\label{eq:OSFTaction}
\end{equation}
where $\Psi$ is a string field of ghost number +1 and $Q_B$ is the BRST charge
\begin{equation}
    Q_B = \frac{1}{2\pi i} \oint \qty( c \, T^{\text{matter}} + : b c \partial c : ) \; 
\end{equation}
with $T^{\text{matter}}$ the total stress-energy tensor of the matter system. The $n$-vertex of the form $\bra{\Psi_1} \ket{\Psi_2 \ast \dots \ast \Psi_n}$ can be defined in terms of a BCFT correlator with $n$ insertions of the corresponding operators $\Psi_i$ on the initial boundary $B_0$ \cite{LeClair:1988sp}. In particular, the 2- and 3-vertices are given by
\begin{align}
    \bra{\Psi_1}\ket{\Psi_2} &= \expval{ I\circ \Psi_1(0) \; \Psi_2(0) }^{\text{UHP}}_{B_0} \label{eq:23vertexi} \\
    \bra{\Psi_1} \ket{\Psi_2 \ast \Psi_3} &= \expval{ f_1 \!\circ\! \Psi_1(0) \; f_2 \!\circ\! \Psi_2(0) \; f_3 \!\circ\! \Psi_3(0) }^{\text{UHP}}_{B_0},
\end{align}
where $f\circ \Psi (z)$ denotes a conformal transformation of the field $\Psi$ by $f$ with
\begin{align}
I(z) &=-\frac{1}{z} \\
f_1(z) &= \tan\qty( \frac{2}{3} \arctan z + \frac{\pi}{3} ) \\
f_2(z) &= \tan\qty( \frac{2}{3} \arctan z ) \\
f_3(z) &= \tan\qty( \frac{2}{3} \arctan z - \frac{\pi}{3} ) .
\label{eq:23vertexf}
\end{align} 

The above action has a large gauge symmetry that needs to be fixed. One possible gauge fixing condition is the Siegel gauge requiring
\begin{equation}
    b_0 \Psi = 0,
\label{eq:Siegel}
\end{equation}
which is particularly suited for level truncation computations and will be used in this note as well. In the Siegel gauge, the BRST charge acting on a string field $\Psi$ of ghost number one has a simple form
\begin{equation}
    Q_B = c_0 L_0^{\text{tot}},
\label{eq:BRSTSiegel}
\end{equation}
where $L_0^{\text{tot}}$ is the sum of $L_0^{\text{matter}}$ and $L_0^{\text{ghost}}$.

Using expressions (\ref{eq:23vertexi}--\ref{eq:23vertexf}) and (\ref{eq:BRSTSiegel}), we can explicitly evaluate the action for any string field in the Siegel gauge. For example, for string fields $\Psi_1=c\phi_1$ and $\Psi_2=c\phi_2$, where $\phi_1$ and $\phi_2$ are primary fields of the matter sector, the kinetic term equals 
\begin{align}
\frac{1}{2} \bra{\Psi_1} \ket{Q_B \Psi_2} 
        &= \frac{1}{2} \bra{\Psi_1} \ket{ c_0 L_0 \Psi_2} = \frac{1}{2} \qty(h_{{\phi}_2}-1) d_{\phi_1\phi_2}g(B_0) \bra{c_{-1}}\ket{c_0 c_1}\\
        &= \frac{1}{2} \qty(h_{{\phi}_2}-1) d_{\phi_1\phi_2} g(B_0),
\label{eq:2vertex}
\end{align}
where $d_{\phi_1\phi_2}$ is the structure constant $C_{\phi_1\phi_2}^{\id}$ and $g(B_0)$ is the $g$-function of the initial boundary, i.e. we have \footnote{For our purposes, it is sometimes useful to consider bases of fields that are not orthonormal with respect to the BPZ product.} $\bra{\phi_1}\ket{\phi_2}=d_{\phi_1\phi_2}g(B_0)$.

Similarly, for three string fields $\Psi_1=c\phi_1$, $\Psi_2=c\phi_2$, $\Psi_3=c\phi_3$, where $\phi_i$ are primaries of conformal weights $h_i$, the 3-vertex equals
\begin{eqnarray}
\bra{\Psi_1} \ket{\Psi_2 \ast \Psi_3} &=& \expval{ f_1 \!\circ\! \Psi_1(0) \; f_2 \!\circ\! \Psi_2(0)\; f_3 \!\circ\! \Psi_3(0) }^{\text{UHP}}_{B_0} \\
   &=& f_{1}'(0)^{h_1-1} f_{2}'(0)^{h_2-1} f_{3}'(0)^{h_3-1} \expval{ \Psi_{1}(\sqrt{3}) \; \Psi_{2}(0) \; \Psi_3(-\sqrt{3}) }^{\text{UHP}}_{B_0}\\
    &=& \qty(\frac{8}{3})^{h_1-1} \qty(\frac{2}{3})^{h_2-1} \qty(\frac{8}{3})^{h_3-1} \expval{ c(\sqrt{3}) \; c(0) \; c(-\sqrt{3}) }^{UHP}\\
    &&\times\expval{ \phi_{1}(\sqrt{3}) \; \phi_{2}(0) \; \phi_3(-\sqrt{3}) }_{B_0}^{UHP} \\
    &=& K^{3-h_1-h_2-h_3}  C_{\phi_1\phi_2\phi_3}g(B_0) ,
\label{eq:3vertex}
\end{eqnarray}
where $K=\frac{3\sqrt{3}}{4}$ and $C_{\phi_1\phi_2\phi_3}=\sum_i d_{\phi_1\phi_i}C^{\phi_i}_{\phi_2\phi_3}$ is a boundary structure constant.

Analogous calculation can be done also for descendant fields. Conveniently, one can use contour-integral deformation to derive so-called conservation laws allowing an iterative calculation of string vertices of descendant fields (see \cite{Rastelli:2000iu} for details).

According to the groundbreaking work of Sen \cite{Sen:1999mh,Sen:1999xm}, nontrivial saddle points of the OSFT action correspond to either the tachyon vacuum (distinguished by having the value of the action equal $-g(B_0)/(2\pi^2)$) or a new OSFT background. Since OSFT backgrounds are associated to a choice of the boundary condition in the bulk CFT, the solutions should be associated to various new conformal boundaries.

Let us now review how to recover some of the properties of the new BCFT from the knowledge of the solution. Sen proposed \cite{Sen:1999mh,Sen:1999xm} that the $g$-function of the background is encoded in the value of the action at the given saddle point
\begin{eqnarray}
g(B_\Psi) =  2 \pi^2 S[\Psi] + g(B_0).
\label{eq:gfunction}
\end{eqnarray}
As proposed in \cite{Kudrna:2012re} (see also \cite{Kiermaier:2008qu} for an alternative method), other coefficients of the boundary state can also be recovered using gauge-invariant quantities called Ellwood invariants \cite{Ellwood:2008jh}, of the form
\begin{equation}
    \bra{E[\mathcal{V}_i]}\ket{\Psi}  = \expval{ \mathcal{V}_i(i,-i) \; f_I\circ \Psi (0)  }^{\text{UHP}}_{B_0},
\label{eq:ellinv}
\end{equation}
where $f_I(z)=2z/(1-z^2)$ and $\mathcal{V}_i$ is a bulk operator of ghost number 2 and conformal weight (0,0). If the matter part of the open string background is factorized\footnote{With an extra non-trivial assumption that the new boundary state can be also written in the factorized form with the $\text{BCFT}^{\text{aux}}$ part unchanged.} into two parts $\text{BCFT}^1\otimes \text{BCFT}^{\text{aux}}$ and the new boundary state of $\text{CFT}^1$ is a linear combination of Ishibashi states
\begin{equation}
\kkett{B_\Psi}^1 = \sum_i n^i_\Psi \: \kett{V_i},
\end{equation}
it was argued by Kudrna, Maccaferri and Schnabl \cite{Kudrna:2012re} that the coefficients $ n^i_\Psi=\bra{V_i}\ket{\ket{B_\Psi}\!\!\:}^1$ are equal to Ellwood invariants
\begin{align}
    n^{i}_{\Psi} = 2\pi i \bra{E[\mathcal{V}_i]}\ket{\Psi-\Psi_{TV}},
\label{eq:Ellwood1}
\end{align}
where $\Psi_{TV}$ is the solution corresponding to the tachyon vacuum, $\mathcal{V}_{i}=c\bar{c}\,V_{i}\otimes\omega_{i}$ with a bulk primary $V_i$ in $\text{CFT}^1$ of conformal weight $(h_i,h_i)$ and an auxiliary bulk primary $\omega_i$ of conformal weight $(1-h_i,1-h_i)$ in $\text{CFT}^{\text{aux}}$ and unit disk one-point function normalized as $\expval{\,\omega_i(0,0)}^{\text{aux}\,}_{\text{disk}}=1$. We refer the reader to \cite{Kudrna:2012re} for a detailed exposition.


\subsection{OSFT and level truncation for defects}
\label{ch:OSFTforDefects}

As discussed in section \ref{subch:defects}, conformal defects in $\text{CFT}$ of central charge $c$ are equivalent to boundaries in $\text{CFT}\otimes \overline{\text{CFT}}$ of conformal charge $2c$ with the folding map reversing the holomorphic and the anti-holomorphic dependence of the folded half.  If we tensor such a folded CFT with an auxiliary $\text{CFT}^{\text{aux}}$ of central charge $c^{\text{aux}}=26-2c$, we get a consistent matter CFT of a bosonic string theory. After tensoring with the $bc$ ghost system, we obtain a CFT of zero total central charge
\begin{equation}
\underbrace{ [ \, \text{CFT} \otimes \overline{\text{CFT}} \, ] }_{ \text{CFT}^{1} } \, \otimes \, \text{CFT}^{\text{aux}}  \, \otimes \, \text{CFT}^{\text{gh}} .
\label{eq:OSFTbackground}
\end{equation}
As the open string background, we choose a boundary state of the above CFT
\begin{equation}
\kkett{B_0} = \kkett{B_0}^{1}  \otimes \kkett{B_0}^{\text{aux}} \otimes \kkett{B_0}^{\text{gh}},
\end{equation}
where $\kkett{B_0}^{1}$ is the boundary state of $\text{CFT}^1=\text{CFT}\otimes\overline{\text{CFT}}$ describing the initial defect after folding. The detailed knowledge of $\kkett{B_0}^{\text{aux}} \otimes \kkett{B_0}^{\text{gh}}$ is not going to be necessary in our discussion since we are going to restrict only to a universal subsector of states associated to the conformal family of the identity.


According to the above discussion, the string field $\Psi$ associated to the saddle point of the action (\ref{eq:OSFTaction}) corresponds to a new OSFT background described by a boundary state $\kkett{B_\Psi}$. Generally, new boundary states might non-trivially mix the three CFT sectors. Since we are interested only in construction of new defects in our CFT, we are going assume that our boundary states factorize into the product of three boundary states with the auxiliary and ghost sector parts unchanged
\begin{equation}
\kkett{B_\Psi} = \kkett{B_\Psi}^{1}  \otimes \kkett{B_0}^{\text{aux}} \otimes \kkett{B_0}^{\text{gh}} .
\end{equation}

The first factor on the right describes a new conformal defect of the unfolded model. In order to identify the defect, one may compute the $g$-function given by (\ref{eq:gfunction}) and other coefficients of the boundary state using Ellwood invariants. The folded theory contains infinitely many primary fields with respect to its total stress-energy tensor. These consist of the tensor products of primary fields on both sides before folding $\phi_\alpha\otimes\phi_\beta$ together with nontrivial combinations of their descendants. The coefficients in front of Ishibashi states associated to tensor products of primary fields 
\begin{equation}
\kkett{B_\Psi}^1 = \sum_{\alpha,\beta} n^{\alpha\beta}_\Psi \: \kett{\phi_\alpha\otimes\phi_\beta} +\dots,
\label{eq:ref}
\end{equation}
with the dots corresponding to Ishibashi states associated to primaries that are not of this simple form, can be easily determined using the Ellwood invariants (\ref{eq:Ellwood1}), i.e.
\begin{align}
    n^{\alpha\beta}_{\Psi} = 2\pi i \bra{E[\mathcal{V}_{\alpha\beta}]}\ket{\Psi-\Psi_{TV}},
\label{eq:Ellwood2}
\end{align}
with $\mathcal{V}_{\alpha\beta}=c\bar{c}\, \phi_\alpha\otimes\phi_\beta\otimes\omega$ and $\omega$ satisfying analogous conditions as above. After unfolding, the coefficient $n_\Psi^{\alpha\beta}$ is equal to the 2-point function of bulk operators $\phi_\alpha$ and $\phi_\beta$ in the presence of the defect
\begin{equation}
    \bra{\phi_\alpha} D_\Psi \ket{\phi_\beta}. 
\end{equation}
Coefficients in front of the other Ishibashi states\footnote{For example, the reflection and transmission coefficients of the defect are encoded in the Ellwood invariants for $((L^{(1)}_{-2}-L^{(2)}_{-2})\otimes (\bar{L}^{(1)}_{-2}-\bar{L}^{(2)}_{-2})\id$, where $L^{(1)}_{-2}$ and $L^{(2)}_{-2}$ together with $\bar{L}^{(1)}_{-2}$ and $\bar{L}^{(2)}_{-2}$ are modes of the stress-energy tensors associated to the two copies of CFT. It is easy to check that this field is a primary field with respect to the total stress-energy tensor with modes $L^{(1)}_{m}+L^{(2)}_{m}$ and $\bar{L}^{(1)}_{m}+\bar{L}^{(2)}_{m}$.} (omitted in the formula (\ref{eq:ref})) can be determined from Ellwood invariants\footnote{Ellwood invariants associated to primaries of high conformal weights turn out to be numerically unstable. For example, the low-level calculation performed in the next section does not give a sensible prediction for reflection and transmission coefficients.} with the use of contour deformations. 

Finding saddle points of the OSFT action is an enormously hard problem and requires an introduction of new nontrivial techniques. The first analytic solution was the tachyon-vacuum solution found by Schnabl in \cite{Schnabl:2005gv} (see also \cite{Okawa:2006vm,Erler:2009uj}). With a sufficient knowledge of the initial BCFT and the new BCFT for which we want to construct the OSFT solution, one can also write a solution in an analytic form as shown by Erler and Maccaferri in \cite{Erler:2014eqa,Erler:2019fye}. Unfortunately, the solution of \cite{Erler:2014eqa,Erler:2019fye} is not going to be useful for our purposes since we want to learn about new consistent defects and thus assuming zero knowledge about the new BCFT associated to our solutions. To accomplish our goal, we have to step back to the numerical method called level truncation. Let us finish this section by reviewing the level truncation extensively used in the SFT literature (see e.g. \cite{Kudrna:2012um, Kudrna:2012re, Kudrna:2016ack, Kudrna:2018mxa, Kudrna:2019xnw, Arroyo:2019liw, AldoArroyo:2011gx, Arroyo:2014pua, Arroyo:2017iis}) and its implementation to explore the space of conformal boundary conditions \cite{Kudrna:2014rya, Vosmera2} and conformal defects.

The space of string fields is given in terms of the boundary spectrum of the above BCFT associated to (\ref{eq:OSFTbackground}) and is badly infinite-dimensional. Out of the all possible fields, one could consistently restrict to subsectors closed under OPEs. All the fields outside of this subsector appear in the action at least quadratically, i.e. at least linearly in the equations of motion, and can be consistently set to zero. Since we are interested in boundary states with the auxiliary sector unchanged, we can restrict to the fields that belong to the identity module of $\text{BCFT}^{\text{aux}}$, i.e. 
\begin{equation}
    \ket{\Psi} = \sum_{i,I,J,K,M} t_{iIJKM} \, L^{(1)}_{-I} L^{(2)}_{-J} \ket{V_i} \otimes L^{\text{aux}}_{-K} \ket{0}^{\text{aux}} \otimes  L^{\text{gh},tw}_{-M}c_1 \ket{0}^{\text{gh}},
\end{equation}
where $V_i$ are defect primaries, $L^{(1)}_{-I}$ and $L^{(2)}_{-J}$ are Virasoro generators on the first and the second part of $\text{CFT}\otimes\overline{\text{CFT}}$, $L^{\text{gh},tw}$ are the twisted\footnote{See e.g. \cite{Kudrna:2019xnw} for the discussion of the necessity of the twisting in the Siegel gauge.} Virasoro generators of the ghost part and $I,J,K,M$ are multi-indices labelling descendants. Note that most of the literature implements also the restriction to the subset of fields that are even under so-called twist symmetry. Since our backgrounds do not admit such a symmetry we will not impose this restriction. On the other hand, as discussed in the next section, we are going to restrict to a particular subsector of fields $\text{CFT}\otimes\overline{\text{CFT}}$ closed under an OPE.

Even after the restriction of the string-field modes from the previous paragraph, the problem of the existence of infinitely many fields persists. This can be solved by approximating the full string field by a truncated space of fields containing only a finite subset of modes. One could then solve the corresponding system of equations of motion, find an approximate solution and check its stability under adding extra fields, i.e. making the approximation more precise. One possible truncation is a restriction to fields with eigenvalue under $L^{\text{tot}}_0+1$ smaller or equal than a fixed level $L$. Generally, one finds a large set of solutions of the truncated system, most of them being a relic of the truncation. To test if a solution is indeed an approximation of a true solution of OSFT, one can check if it is stable under the increase of the level. In the usual implementation of the method, one solves the truncated equations at low levels and looks for solutions at higher levels using the Newton's method starting with an approximation given by the lower-level solution. If the Newton's method converges to a nice solution with sensible value of the $g$-function that is close to the initial point, it suggests a stability of the solution. Obviously, it is unlikely that all solutions of OSFT can be constructed using such a method but as illustrated bellow, it indeed leads to many interesting solutions.

\section{Conformal defects from OSFT}
\label{ch:Results}

In this section, we apply the OSFT approach outlined above to find solutions corresponding to the fixed points of KRW \cite{Kormos:2009sk}.

Since general (large $m$) minimal models contain large number of fields already at low levels, we need to perform a further restriction of string field components. A natural truncation is a restriction to the subsector of fields that are closed under the OPE of $\phi_{(1,3)(1,1)}$ and $\phi_{(1,1)(1,3)}$. This indeed leads to a nice family of operators with a stable (finite) amount of fields of conformal weight smaller than level $L$ even in the large $m$ limit. The reason is the following. Representations with Kac labels $(1,2k+1)$ are closed under fusion. Moreover, out of all the combinations of such representations $R_{(1,2k+1)}\otimes \bar{R}_{(1,2l+1)}$, only a special class appears in the spectrum of our defect. In particular, there are pairs of fields with labels $\phi_{(1,2k-1)(1,2k-1)}$ for integral parameter $0<k< \frac{m+1}{2}$ together with fields $\phi_{(1,2k+1)(1,2k-1)}$ and $\phi_{(1,2k-1)(1,2k+1)}$ for each integral $0<k< \frac{m-1}{2}$. The conformal weight of such fields increases quadratically with $k$ and as a result there is always only a finite number of primary fields up to a given level.

In order to compute the interaction part of the action, i.e. the 3-vertices (\ref{eq:3vertex}), one needs to know the defect structure constants. These have been derived for fields $\phi_{(1,3)(1,1)}, \phi_{(1,1)(1,3)}$ together with the two copies\footnote{Note that the Ising model contains only a single defect field $\phi_{(1,3)(1,3)}$ with structure constants reviewed in the appendix \ref{app:StructureConstants} as well.} of $\phi_{(1,3)(1,3)}$ for $m>3$ in \cite{Makabe:2017uch}. For the convenience of the reader, we list them in appendix \ref{app:StructureConstants}. The OPEs of $\phi_{(1,3)(1,1)}$ and $\phi_{(1,1)(1,3)}$ for $m\geq 5$ include additional fields $\phi_{(1,5)(1,3)}$ and $\phi_{(1,3)(1,5)}$ of conformal weight $h=5-8/(m+1)$ whose structure constants has not been derived. Therefore, in the following tables and plots, we always restrict to computations up to levels:
\begin{alignat}{2}
    &L=6 \quad &&\text{for} \quad m=3,4 \nonumber \\
    &L=3.5 \quad &&\text{for} \quad m=5,6 \label{eq:levels} \\
    &L=4 \quad &&\text{for} \quad m\geq 7 . \nonumber
\end{alignat}
In particular, we know all the necessary structure constants for the $m=3,4$ cases and the only limit are the numerical issues preventing us from dealing with too many fields. For $m=5,6$, the field $\phi_{(1,5)(1,3)}$ has dimension lower than $4$ and the corresponding structure constants are needed for the treatment of the models at level 4. 

We compute the $g$-functions of numerical solutions from the action using formula (\ref{eq:gfunction}). For each $m$, we extrapolate the $g$-function of an OSFT solution to an infinite level by fitting a general quadratic function in $1/L$ using approximations at all integral levels and sending $L\rightarrow\infty$. A comparison of the fits and solutions at different levels of truncation is shown in figures \ref{fig:comparefitC} and \ref{fig:comparefitIsing1}. The values of the $g$-functions for our solutions at integral levels, together with the extrapolated values and the predictions from the literature are shown in figures \ref{fig:extrapolations} and \ref{fig:comparefitIsing1}.

The Ellwood invariants (\ref{eq:ellinv}) computed for numerical solutions are known to have oscillatory behavior (with the period of four levels) that increases with the conformal weight of the corresponding state whose coefficient we want to compute \cite{Kudrna:2019xnw}. These oscillations limit us from studying boundary-state coefficients using low-level computations. Therefore, we include an Ellwood invariant calculation only for the Ising model where level 6 computation is possible and where the calculation is necessary to distinguish between the continuum of found solutions. 

Let us now concentrate on the analysis of the Ising model and $m>3$ minimal models in greater detail.

\subsection{$m>3$ minimal models}

Let us start with the analysis of the string field truncated to $L=1$ for $m>3$. The truncated field has the form
\begin{equation}
    \ket{\Psi_{L=1}} = \qty( t_1\ket{0}+ t_2 \, \ket{\phi_{(1,3)(1,1)}} + t_3 \ket{\phi_{(1,1)(1,3)}} ) \otimes \ket{0}^{\text{aux}} \otimes c_1\ket{0}^{\text{gh}},
    \label{eq:stringfield1}
\end{equation}
where $t_i$ are the string field coefficients. Plugging into the prescription for the action (\ref{eq:2vertex}) and (\ref{eq:3vertex}), we get
\begin{align}
S[\Psi_{L=1}] =& -\frac{g_0}{2} \qty( -t_1^2 - (t_2^2+t_3^2) (h-1) d_{\phi\phi} ) \\ 
&-\frac{g_0}{3} \qty( t_1^3K^3 + (t_2^3+t_3^3)K^{3-3h} d_{\phi\phi}C_{\phi\phi}^{\phi} + 3(t_1t_2^2+t_1t_3^2) K^{3-2h} d_{\phi\phi} ) ,
\end{align}
where $K=3\sqrt{3}/4$, $h=(m\!-\!1)/(m\!+\!1)$ is the conformal dimension of $\phi_{(1,3)(1,1)}$ and $\phi_{(1,1)(1,3)}$ and $d_{\phi\phi}$ and $C_{\phi\phi}^{\phi}$ are the structure constants listed in the appendix \ref{app:StructureConstants}.

\begin{figure}[t]
\centering
\begin{subfigure}[t]{0.48\textwidth}
    \centering
	\includegraphics[width=\textwidth]{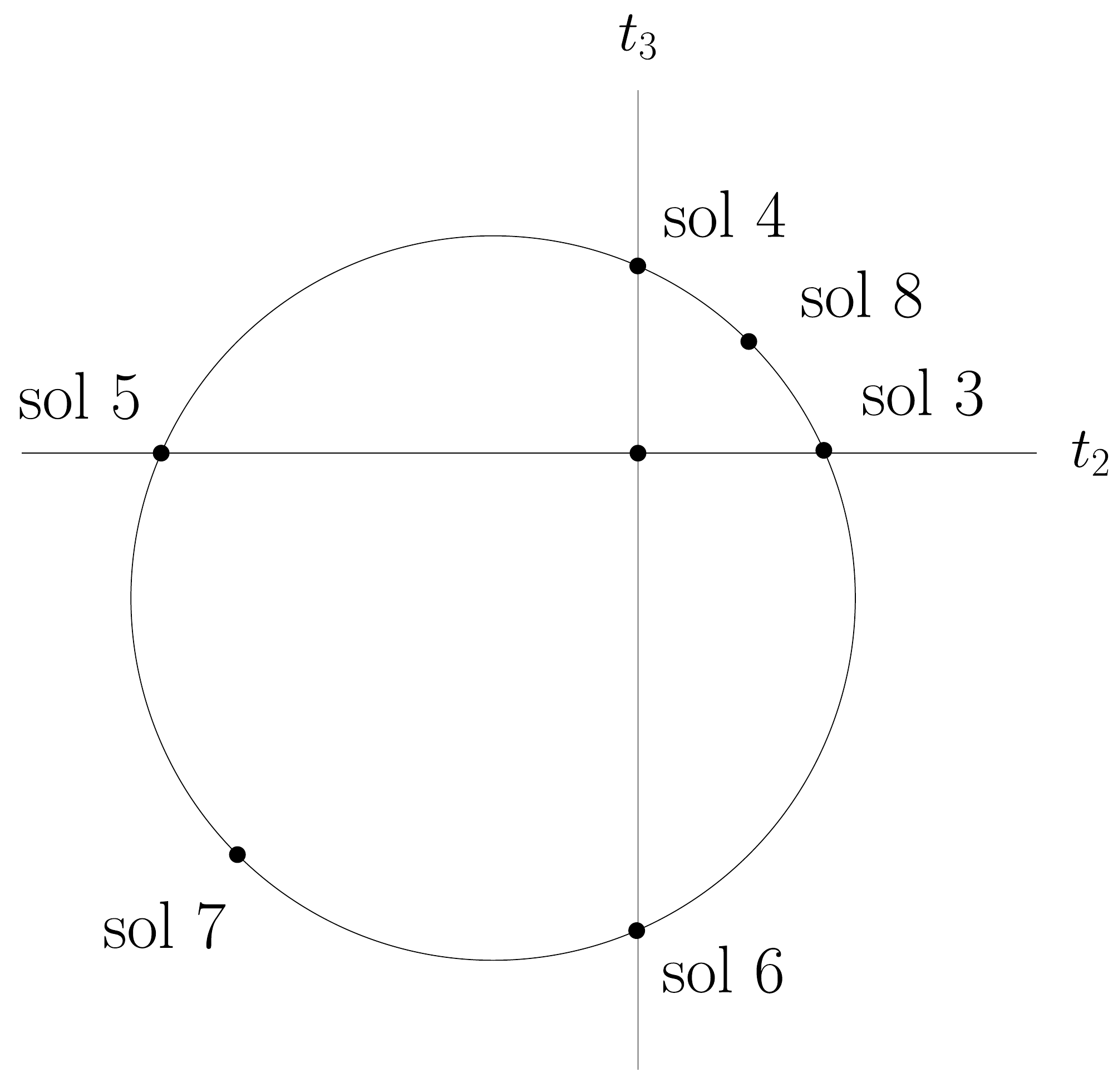}
	\caption{OSFT solutions from table \ref{tab:largem-lev1}.}
	\label{fig:osftsols}
\end{subfigure}
\hspace{0.1cm}
\begin{subfigure}[t]{0.48\textwidth}
    \centering
	\includegraphics[width=\textwidth]{images/Largem.pdf}
    \caption{RG fixed points reproduced from \cite{Kormos:2009sk}. }
    \label{fig:largem-flows2}
\end{subfigure}
\caption{Structure of OSFT solutions at $L=1$ matches the structure of RG fixed points found by KRW.}
\label{fig:osft-largem}
\end{figure}

It is a simple task to search for saddle points of the action and one obtains 8 solutions listed in table \ref{tab:largem-lev1}. The list of solutions contains the obvious trivial solution associated to the original defect, together with the tachyon vacuum solution always present in any background and corresponding physically to the disappearance of open string modes (no boundary). Looking at the $t_2-t_3$ plane corresponding to the fields $\phi_{(1,3)(1,1)}, \phi_{(1,1)(1,3)}$ and disregarding the tachyon vacuum solution, we recover the same structure of fixed points as in the RG flow analysis of \cite{Kormos:2009sk} illustrated in figure \ref{fig:osft-largem}. Note that the full picture with six IR fixed points is recovered already at level one compared to the leading order conformal perturbation theory that leads to only three fixed points. 

\begin{table}[t]
\centering
\begin{tabular}{|c|l|lll|c|}
\hline
nr & $g$-function & solution & & & defect \\ \hline
1 & 1.91899 & $t_1\to 0$, & $t_2\to 0$, & $t_3\to 0$ & $D_{(1,2)}$ \\ 
2 & 0.60522 & $t_1 \to 0.456178$, & $t_2\to 0$, & $t_3\to 0$ & TV \\ 
3 & 1.90717 & $t_1\to 0.0136518$, & $t_2\to 0.108315$, & $t_3\to 0$ &  $D_{(2,1)}$ \\ 
4 & 1.90717 & $t_1\to 0.0136518$, & $t_2\to 0$, & $t_3\to 0.108315 $& $D_{(2,1)}$ \\ 
5 & 1.54999 & $t_1\to 0.209826$, & $t_2\to -0.316835$, & $t_3\to 0$ & $D_{(1,1)}$ \\ 
6 & 1.54999 & $t_1\to 0.209826$, & $t_2\to 0$, & $t_3\to -0.316835$ & $D_{(1,1)}$ \\ 
7 & 1.66701 & $t_1\to 0.163031$, & $t_2\to -0.215421$, & $t_3\to -0.215421$ & $F$ \\ 
8 & 1.90052 & $t_1\to 0.0205803$, & $t_2\to 0.0932996$, & $t_3\to 0.0932996$ & $C$ \\ 
\hline
\end{tabular}
\caption{Eight solutions at the truncation level $L=1$ for $m=10$. The last column shows proposed identification of solutions with conformal defects (‘‘TV'' denotes the tachyon vacuum).}
\label{tab:largem-lev1}
\end{table}

\begin{figure}[h]
    \centering
    \begin{subfigure}[c]{\textwidth}
    \centering
    \includegraphics{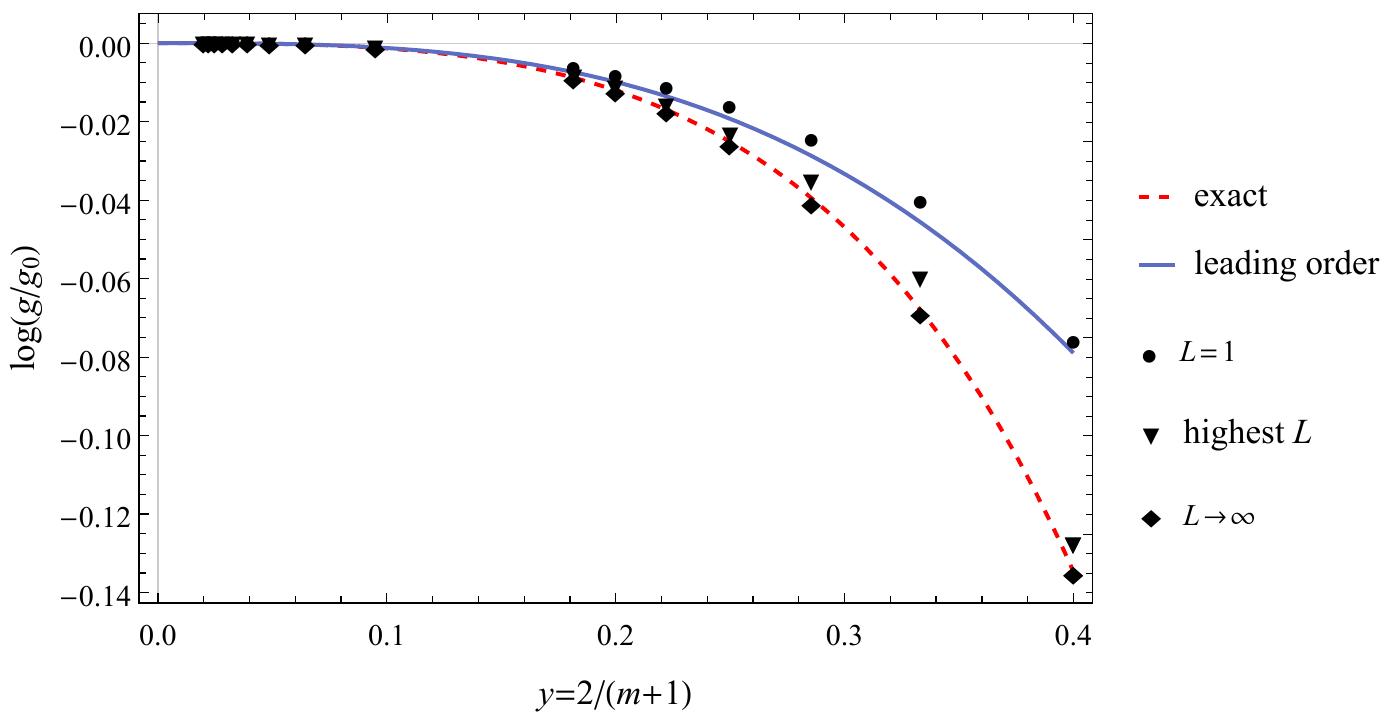}
    \caption{Comparison of the exact $g$-function of defect $D_{(2,1)}$, the leading order $g$-function of the fixed point (\ref{eq:gD21pert}) and $g$-functions of OSFT solutions at the highest studied levels (\ref{eq:levels}), as well as their extrapolations to $L\rightarrow \infty$. }
    \label{fig:extrapolationD}
    \end{subfigure}
    \bigskip
    \medskip
    
    \begin{subfigure}[c]{\textwidth}
    \centering
    \includegraphics{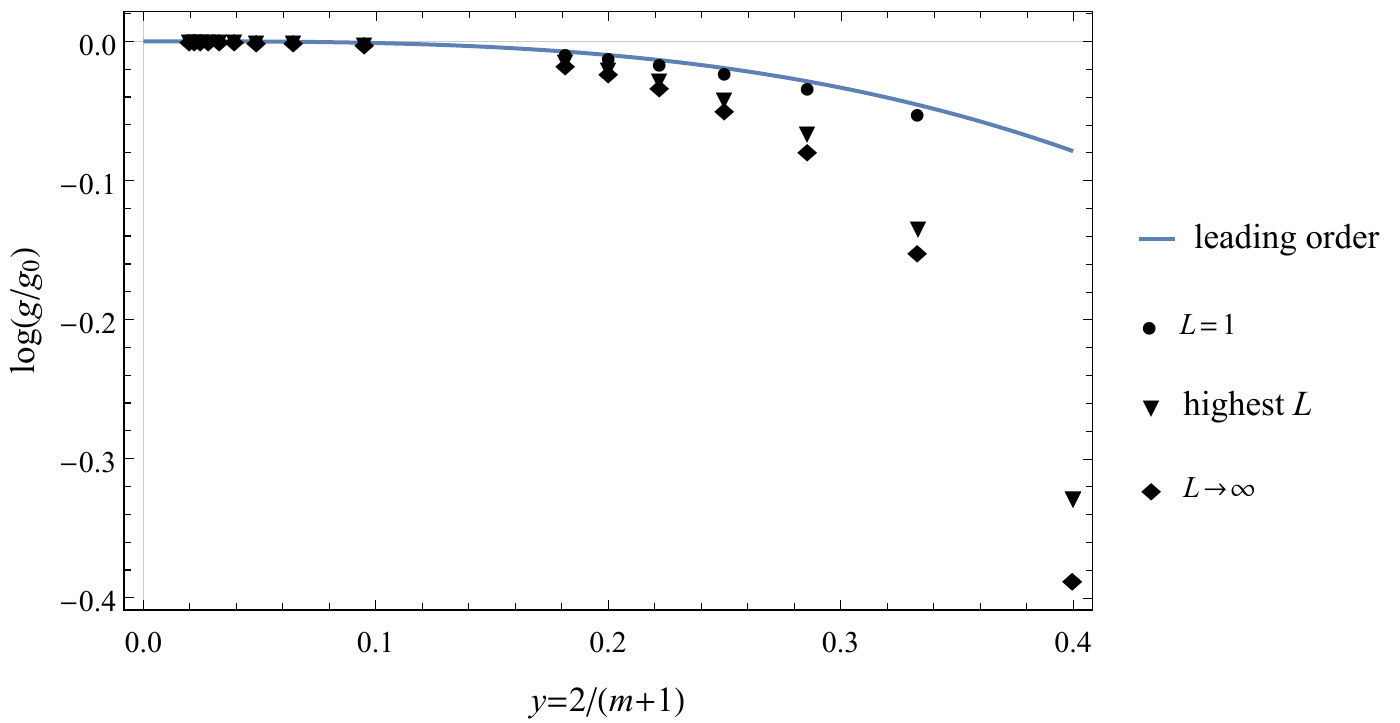}
    \caption{Comparison of the leading order $g$-function of the fixed point (\ref{eq:gCpert}) corresponding to defect $C$ and $g$-functions of OSFT solutions at the highest studied levels (\ref{eq:levels}), as well as their extrapolations to $L\rightarrow \infty$.}
    \label{fig:extrapolationC}
    \end{subfigure}
\caption{Leading order $g$-functions of fixed points corresponding to defects $D_{(2,1)}$ and $C$ compared to OSFT values. In figure (\ref{fig:extrapolationD}), the OSFT $g$-functions fit better the exact $g$-function of defect $D_{(2,1)}$. We expect the OSFT values in figure (\ref{fig:extrapolationC}) to give a better prediction of the $g$-function of defect $C$ then the leading-order conformal-perturbation-theory result.}
\label{fig:extrapolations}
\end{figure}

From now on, let us focus on the four OSFT solutions that correspond to distinct defects\footnote{Note that there is the reflection symmetry of the picture that switches the solutions 3 and 4 and solutions 6 and 5. The corresponding coefficients of the string field differ by a sign and we can only restrict to one of the solutions. We will see that the reflection symmetry in the $m=3$ case can be identified with the orbifold $\mathbb{Z}_2$ action of the description in terms of the free boson on $S^1/\mathbb{Z}_2$.} $D_{(2,1)}$, $D_{(1,1)}$, $F$ and $C$. To test if the above solutions are indeed approximations of a true OSFT solution and to check the match of the $g$-function, one can use the Newton's method to find a more accurate approximation at level $L+0.5$, starting from the solution at level $L$ (and setting coefficients of the newly added fields to zero for the approximation used in the Newton's method). We compute the solutions using the Newton's method up to levels indicated in (\ref{eq:levels}). We plot the $1/m$ dependence of $g$-functions of the OSFT solutions at different levels of truncation in figures \ref{fig:plotD21}--\ref{fig:plotF}. For defects $D_{(2,1)}$, $D_{(1,1)}$ and $F$, we compare them with the exact $g$-functions. In the case of defects $D_{(2,1)}$ and $C$, we can also compare our results with the prediction coming from the leading order conformal perturbation theory
\begin{align}
\log g(D_{(2,1)}) -\log g_0 &= - \frac{\pi^2}{8} y^3 + \mathcal{O}(y^4) \label{eq:gD21pert} \\
\log g(C) - \log g_0 &= - \frac{\pi^2}{4} y^3 + \mathcal{O}(y^4), \quad y=\frac{2}{m+1} \label{eq:gCpert}
\end{align}
shown in figures \ref{fig:extrapolationD} and \ref{fig:extrapolationC}. Note that level-one OSFT predictions for the $g$-functions of defects $D_{(2,1)}$ and $C$ match well the leading order conformal perturbation theory. On the other hand, one can see that the level-one predictions for the $g$-functions of the identity defect and defect $F$ are not even qualitatively correct (see e.g. the non-zero slope for the $D_{(1,1)}$ defect). As the level increases, we can see a convergence to the expected value with the extrapolated values matching the exact values reasonably well.
 
\begin{figure}[t]
      \centering
      \subcaptionbox{Fit for integer levels. \label{fig:comparefitC1}}
        {\includegraphics[scale=1]{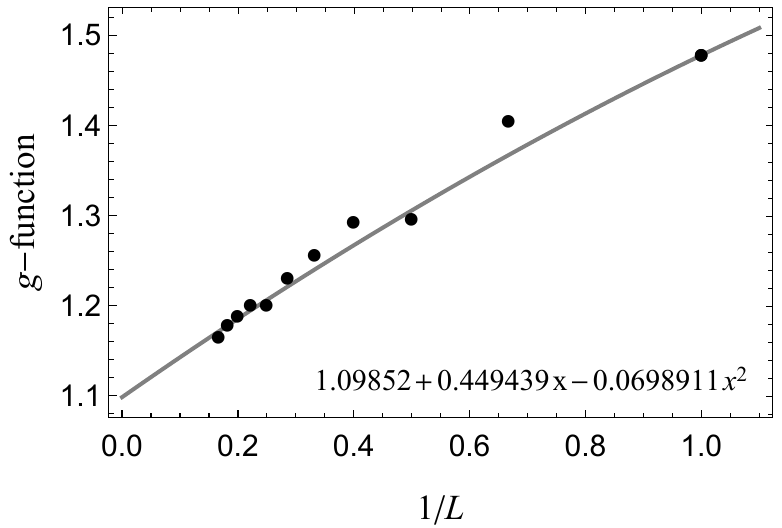}}
      \subcaptionbox{Fits for even, odd and (a part of) half-integer levels.\label{fig:comparefitC3}}
        {\includegraphics[scale=1]{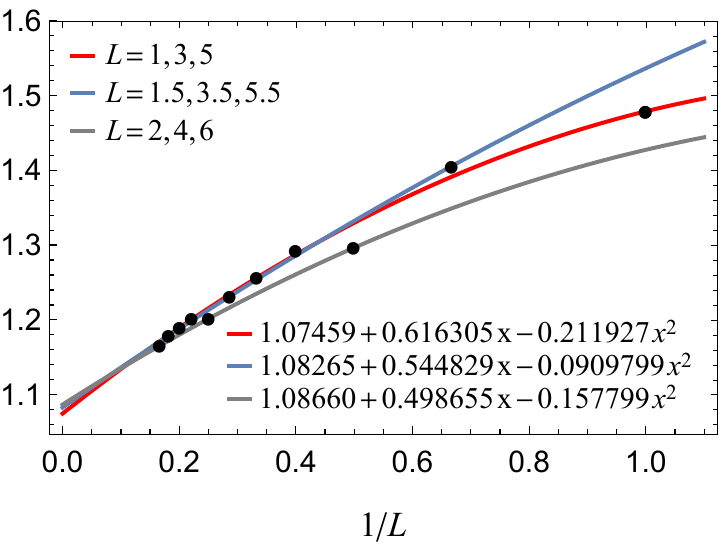}}
      \caption{Comparison of quadratic fits and $g$-functions of numerical OSFT solutions for the defect $C$ in the tricritical Ising model $m=4$. }\label{fig:comparefitC}
\end{figure}

By looking at the $g$-functions of the defect $D_{(2,1)}$, we observe that OSFT solutions at low levels together with the extrapolation to the infinite level give a prediction that is much closer to the exact value than the leading-order perturbative $g$-function (\ref{eq:gD21pert}) for small $m$. We therefore expect that our OSFT computation gives a better estimate for the $g$-function of the defect $C$ for small $m$. The $g$-functions\footnote{The Newton's method fails in finding the level 1.5 solution starting from the level 1 approximation for $m=4$. The solution at level 1.5 was found by exactly solving the equations for saddle points.} of defect $C$ predicted by OSFT are presented\footnote{The $g$-functions of the four nontrivial solutions are listed in appendix \ref{app:Tables}.} in table \ref{tab:gC} in appendix \ref{app:Tables}. In particular, the estimated value for the case of the tricritical Ising model $m=4$ is around $g_C\approx 1.099$. The $g$-functions of numerical OSFT solutions exhibit oscillatory behavior with the period of two levels. Therefore, in case of the tricritical Ising model, where level 6 computations are possible, we can try to extrapolate to infinite level separately for even, odd and half-integral levels \footnote{Because of the aforementioned periodicity of two levels, in figure \ref{fig:comparefitC3} we fit to levels $L=1.5, \, 3.5, \, 5.5$ but not to $L=2.5, \, 4.5$.} (instead of fitting to all integral levels like in table \ref{tab:gC}). We show the different fits and their comparison to the numerical values in figure \ref{fig:comparefitC}. The average of the extrapolations from figure \ref{fig:comparefitC3} suggests that a better estimate for the $g$-function of the defect $C$ for the tricritical Ising model is in fact closer to
\begin{align}
    g_{\text{OSFT}}(C) \approx 1.081.
    \label{eq:gCm4}
\end{align}
Higher-level calculations are necessary for a more accurate prediction. 

\begin{figure}[h]
    \centering
\begin{subfigure}[c]{\textwidth}
    \centering
    \includegraphics{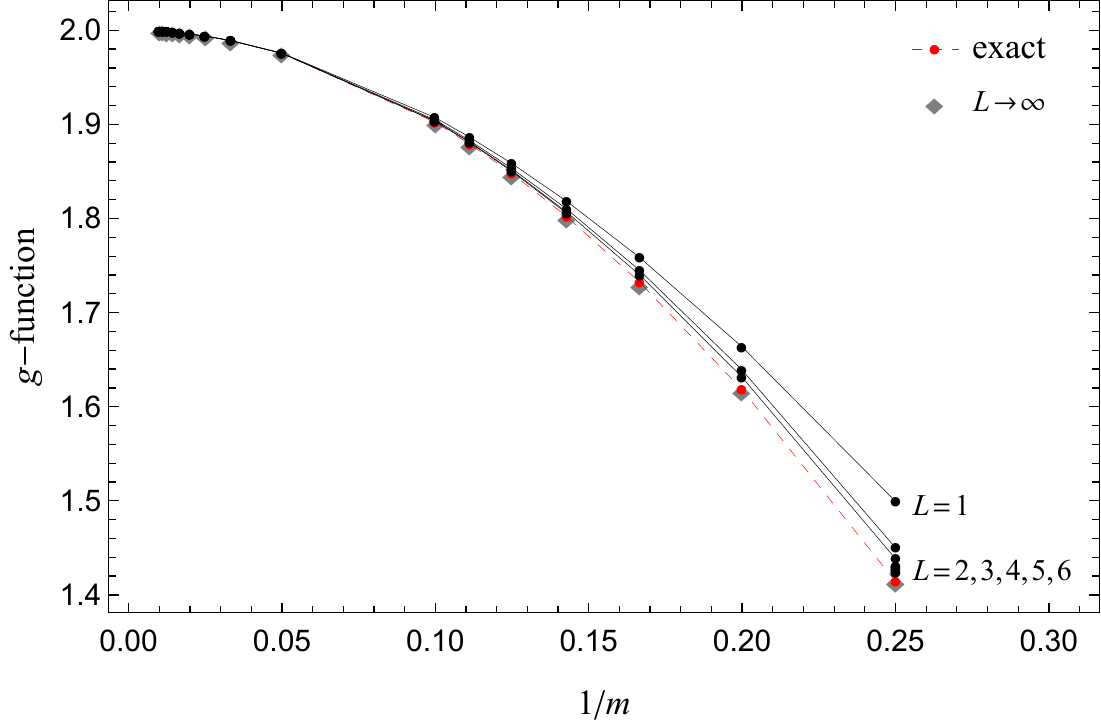}
    \caption{The $g$-function of the third OSFT solution from table \ref{tab:largem-lev1} at different levels of truncation compared to the exact $g$-function of defect $D_{(2,1)}$.}
    \label{fig:plotD21}
\end{subfigure}
\bigskip
\medskip

\begin{subfigure}[c]{\textwidth}
    \centering
    \includegraphics[width=0.7\textwidth]{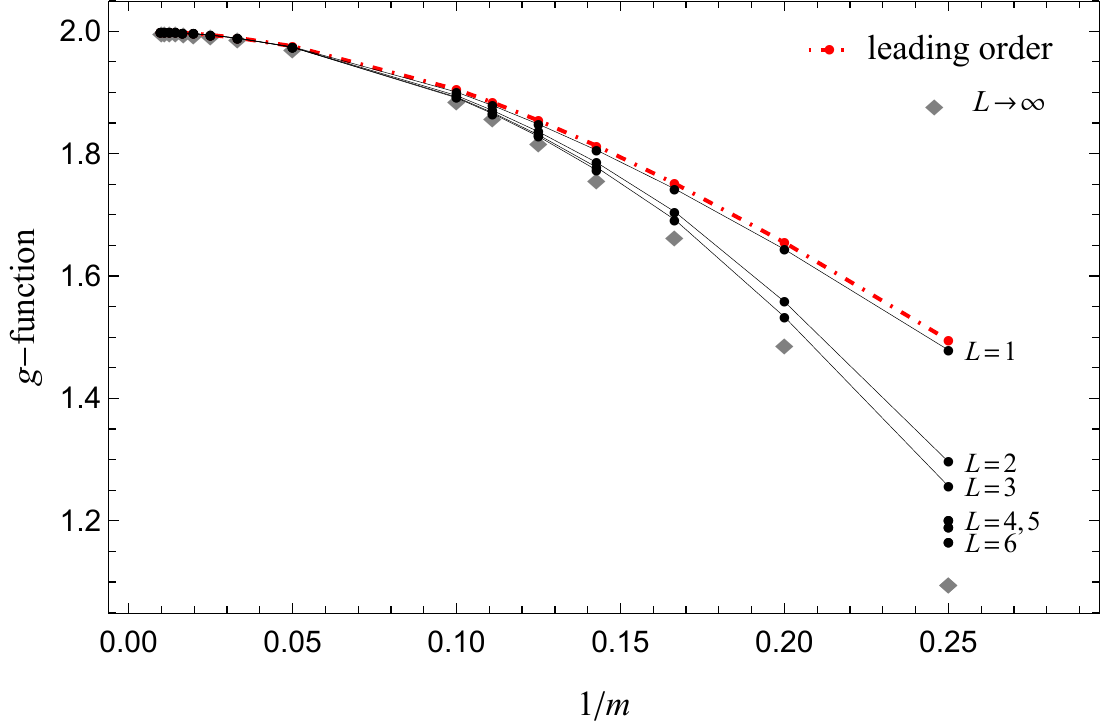}
    \caption{The $g$-function of the last OSFT solution from table \ref{tab:largem-lev1} at different levels of truncation compared to the leading order $g$-function of defect $C$.}
    \label{fig:plotC}
\end{subfigure}
\end{figure}
\begin{figure}[h]\ContinuedFloat
\centering
\begin{subfigure}[c]{\textwidth}
    \centering
    \includegraphics[width=0.7\textwidth]{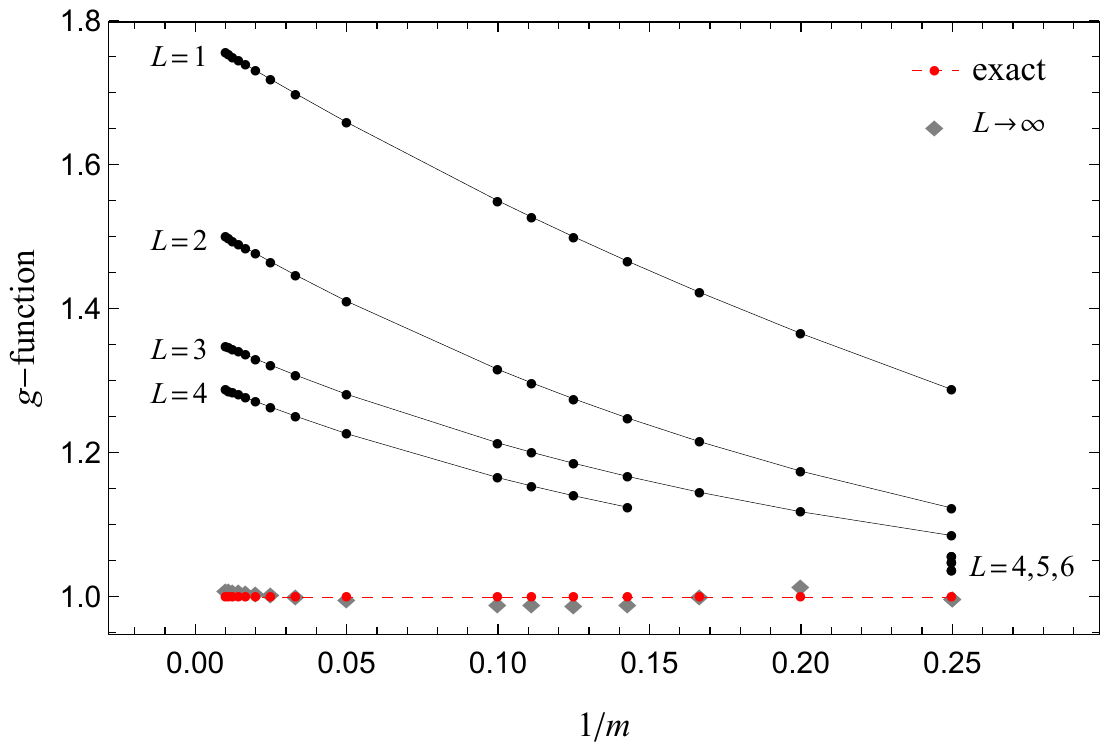}
    \caption{The $g$-function of the fifth OSFT solution from table \ref{tab:largem-lev1} at different levels of truncation compared to the exact $g$-function of the trivial defect $D_{(1,1)}$.}
    \label{fig:plotD11}
\end{subfigure}
\bigskip
\medskip

\begin{subfigure}[c]{\textwidth}
    \centering
    \includegraphics[width=0.7\textwidth]{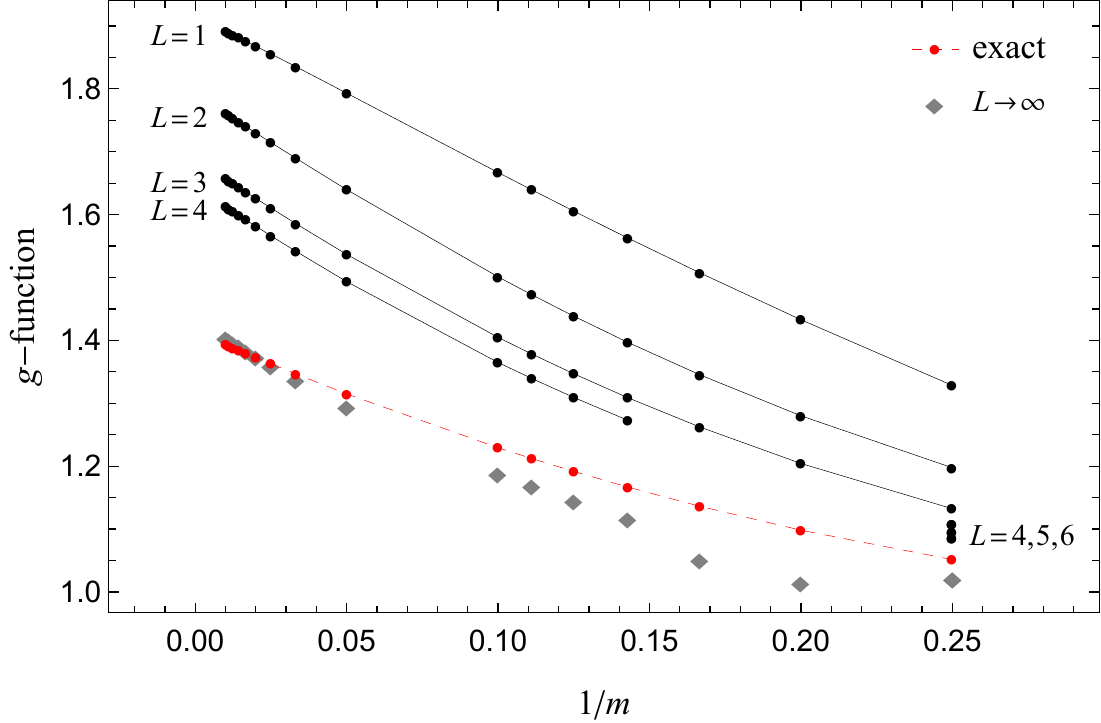}
    \caption{The $g$-function of the seventh OSFT solution from table \ref{tab:largem-lev1} at different levels of truncation compared to the exact $g$-function of the defect $F$ as defined in (\ref{eq:defectF}).}
    \label{fig:plotF}
\end{subfigure}
\caption{$g$-functions of OSFT solutions listed in (\ref{eq:levels}) compared to exact or leading order $g$-functions for defects $D_{(2,1)}$, $C$, $F$ and $D_{(1,1)}$. }
\label{fig:gvalues}
\end{figure}
\FloatBarrier

\subsection{Ising model}

\begin{figure}[t]
      \centering
      \includegraphics[scale=1]{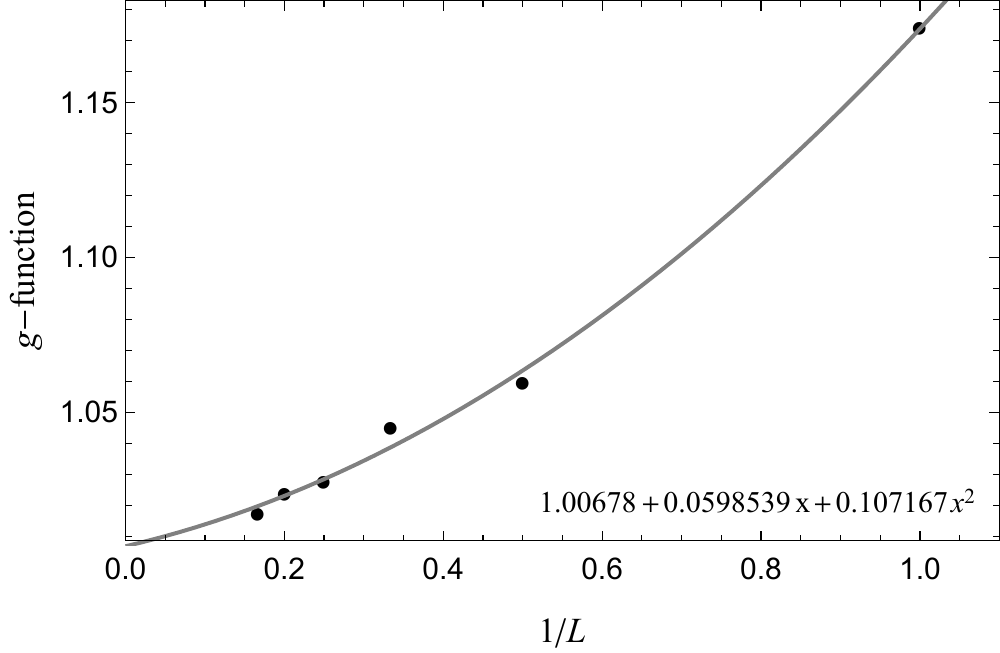}
      \caption{Dependence of the value of the $g$-function on the inverse level for the continuum of solutions in the Ising model. The extrapolated value of the $g$-function equals $1.007$.}
      \label{fig:comparefitIsing1}
\end{figure}

In the case of the Ising model, one more field $\phi_{(1,3)(1,3)}$ is present already at level $L=1$. Let us thus start with the analysis up to level $L=0.5$ with the string field as in (\ref{eq:stringfield1}) and the action
\begin{equation}
  S[\Psi_{L=1/2}]=-\frac{1}{\sqrt{2}}(-t_1^2 -\frac{1}{2} d_{\phi\phi} (t_2^2+t_3^2) ) -\frac{\sqrt{2}}{3} (K^{3} t_1^3 + 3d_{\phi\phi} K^{2} t_1(t_2^2+t_3^2)),
\end{equation}
where $K=3\sqrt{3}/4$. Note that the action is (at least up to this low level) invariant under the rotation in the $t_2-t_3$ plane, leading to a continuum of solutions (together with the trivial solution and the tachyon vacuum) with the value of the $g$-function $1.1742$ not far from the expected value $g=1$. An analysis at higher levels reveals a continuum of solutions as well, with equal values of the $g$-function shown in figure \ref{fig:comparefitIsing1}. The quadratic extrapolation to the infinite level gives value $g=1.007$ in a remarkable agreement with the expected value $g=1$.

We would like to identify solutions parametrized by $t_3/t_2$ with continuum of defects parametrized by $\lambda_r/\lambda_r$ or equivalently by $\tan \alpha=\lambda_r/\lambda_r$. To distinguish between the solutions, one needs to determine some of the coefficients in front of the Ishibashi states of the boundary state in the folded Ising model
\begin{equation}
    \kkett{B_\Psi} = \sum_{(r,s),(x,y)\in\mathcal{I}_3} n_\Psi^{(r,s)(x,y)} \kett{\phi_{(r,s)} \otimes \phi_{(x,y)}}+\dots .
    \label{eq:ref2}
\end{equation}
From the perspective of the orbifolded free boson, the continuum of boundary states with $g$-functions equal one are D0-branes located at a given position on the circle. The position can be obviously distinguished by the expectation value of the vertex operator $\cos 2\phi$ at the boundary. The value of the field $\phi$ at the Dirichlet boundary $D(\varphi)$ is fixed to $\varphi$ with the expectation value of $\cos 2\phi$ being $\cos 2\varphi$. The vertex operator $\cos 2\phi$ can be identified with the bulk operator $\frac{1}{2}(\epsilon \otimes \id+ \id \otimes \epsilon)$ in the folded model with the expectation value encoded by 
\begin{align}
    \frac{1}{2} \qty(n_\Psi^{(1,3)(1,1)}+n_\Psi^{(1,1)(1,3)} ).
\end{align}

As explained in section \ref{ch:OSFTforDefects}, the coefficient $n_\Psi^{(1,3)(1,1)}$ is given by the Ellwood invariant with the insertion of the bulk field $\phi_{(1,3)}\otimes\phi_{(1,1)}=\varepsilon\otimes\id$:
\begin{align}
n_{\Psi}^{(1,3)(1,1)} &= 2\pi i \bra{E[\mathcal{V}_{(1,3)(1,1)}]}\ket{\Psi-\Psi_{TV}} \\
&= 2\pi i \bra{E[c\bar{c}\,\varepsilon\otimes\id\otimes\omega]}\ket{\Psi-\Psi_{TV}},
\label{eq:n1311}
\end{align}
where $\omega$ is an auxiliary field of conformal weight $(1-h_{1,3},1-h_{1,3})=(1/2,1/2)$ and one-point function
\begin{align}
\expval{\,\omega(i,-i)}^{\text{aux}\,}_{\text{UHP}}= 4^{h_{1,3}-1}\expval{\,\omega(0,0)}^{\text{aux}\,}_{\text{disk}}=\frac{1}{\sqrt{2}}  .  
\end{align}
When computing Ellwood invariants, the tachyon vacuum solution can be substituted with $\frac{2}{\pi}c_1\ket{0}$ and therefore $\Psi_{TV}$ does not contribute to the above Ellwood invariant \cite{Kudrna:2012re}. 

\begin{figure}[t]
	\centering
	\includegraphics{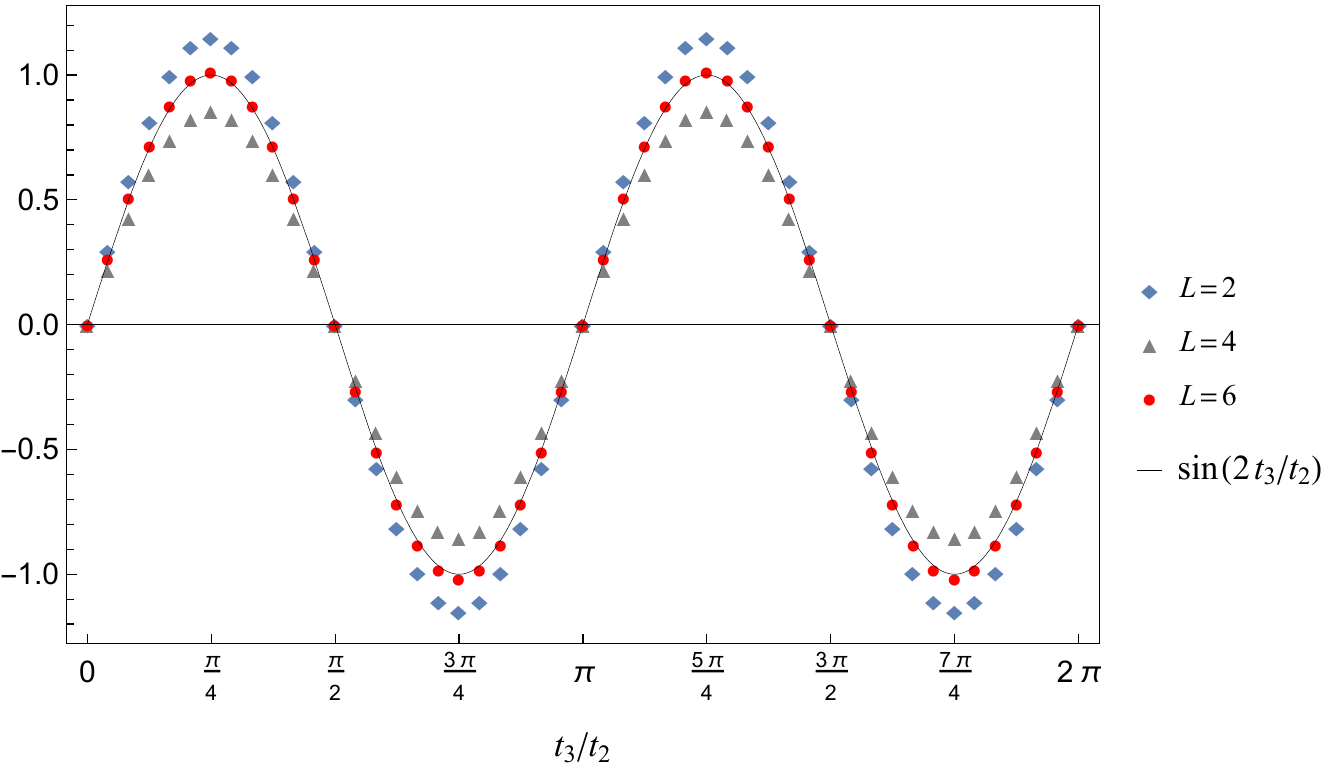}
	\caption{The value of the Ellwood invariant associated to $(\varepsilon\otimes\id+\id\otimes\varepsilon)/2$ at levels 2, 4 and 6 together with the expected value depending on the ratio $\tan \alpha= t_3/t_2$ of the string field coefficients appearing in front of the fields $\phi_{(1,3)(1,1)}$ and $\phi_{(1,1)(1,3)}$. The numerical values have a correct oscillatory behavior and roughly match the expected value. The oscillation with respect to the level does not allow us to perform a confident extrapolation as in the case of the $g$-function.}
	\label{fig:Ising_Ellwood_Invariants}
\end{figure}

The prediction coming from the Ellwood invariants\footnote{Note the Ellwood invariant associated to the difference $\varepsilon\otimes\id-\id\otimes\varepsilon$ vanishes as expected.} is shown in the figure \ref{fig:Ising_Ellwood_Invariants} for levels $2,4$ and $6$ and allows us to identify the ratio $t_3/t_2$ with the ratio $\lambda_r/\lambda_l$ from \cite{Kormos:2009sk}. Note that (as expected) the convergence of Ellwood invariants is worse that the one of the $g$-function. As a result, we are not able to find a sensible extrapolated values using the limited number of points. We can still see a good agreement with the expected value. 

\section{Discussion}
\label{ch:Discussion}

In this note, we described a method of finding conformal defects with the help of the level-truncation technique in OSFT. Using the folding trick \cite{Wong:1994np}, we converted the problem of finding conformal defects to the problem of finding conformal boundaries. A conformal boundary can be interpreted (after tensoring with auxiliary sectors) as an OSFT background. Starting from an initial background, one can search for the solutions to OSFT equations of motion, conjecturally corresponding to new open-string backgrounds, and interpret them as new boundaries of the world-sheet CFT. Level truncation is a powerful numerical method that allows an exploration of the space of such solutions. From numerical OSFT solutions, one can then extract estimates of various quantities associated to the new boundary, such as the $g$-function and other coefficients of the corresponding boundary state using Ellwood invariants \cite{Kudrna:2012re}.

We illustrate the method by looking for solutions around the background descending from the  $D_{(1,2)}$ topological defect in minimal models $\mathcal{M}_{m+1,m}$. Already at level $L=1$ in the level truncation method, we recover the structure of fixed points from \cite{Kormos:2009sk} found using the combination of conformal perturbation theory and the truncated conformal space approach. We compute a numerical prediction for $g$-functions of selected solutions at low levels and find  their extrapolated value at infinite level. The accuracy of the level-one prediction is comparable to the leading-order conformal perturbation theory with higher-level results providing much better accuracy. The extrapolated values of the $g$-function for previously identified defects match the expected values reasonably well.

Just like in \cite{Kormos:2009sk}, we find a numerical evidence for the existence of the conformal defect $C$ in $m\geq 4$ minimal models\footnote{The would-be defect $C$ in the Ising model is simply $\kkett{\id}\!\!\bbraa{\varepsilon} + \kkett{\varepsilon}\!\!\bbraa{\id}$.} and give a numerical prediction for its $g$-function dependence on $m$. The TCSA method in \cite{Kormos:2009sk} was inconclusive in case of the tricritical Ising model with $m=4$. In our analysis, we experienced an issue with convergence of the Newton's method at level $L=1.5$ when starting from the level $L=1$ approximation. Finding the approximated solution for the defect $C$ directly at level 1.5 and using this point as an approximation for the higher-level analysis leads to a well-behaved solution. It is tempting to speculate that the truncated-conformal-space-approach issues are related to our numerical difficulties at low levels. Luckily, the fields $\phi_{(1,5)(1,3)}$ and $\phi_{(1,3)(1,5)}$ are not present in case of the tricritical Ising model and we are able to study the solution up to level $L=6$ with our current Mathematica code and get the prediction $g_C=1.081$ for its $g$-function. Higher-level computations are necessary in order to determine other boundary-state coefficients of this defect by computing Ellwood invariants.

The case of the Ising model with $m=3$ is qualitatively different from other minimal models. We show that OSFT equations of motion allow continuum of solutions (at least up to level $6$) that are in correspondence with the continuum of expected RG fixed points  of \cite{Kormos:2009sk, Fendley_2009}. We find a nice match in the value of the $g$-function and compute the combination of the overlaps $\qty(\bra{\varepsilon} D \ket{\id} + \bra{\id} D \ket{\varepsilon})/2$ using Elwood invariants that allows us to differentiate the corresponding defects.

In this note, we identified string field solutions with conformal defects by matching the structure of the space of solutions to the structure of RG fixed points from \cite{Kormos:2009sk} and comparing their $g$-functions. Ideally, we would like to know more information about obtained conformal defects, such as other overlaps $\bra{\phi_\alpha} D \ket{\phi_\beta}$ or the reflection and transmission coefficients \cite{Quella:2006de, Kimura:2014hva}. These quantities can be in principle determined from numerical OSFT solutions using Ellwood invariants by the prescription of \cite{Kudrna:2012re} as we did in the example of the Ising model. However, the calculation of boundary-state coefficients requires the knowledge of the bulk-to-boundary structure constants for the original defect. Furthermore, values computed from numerical solutions have oscillatory behavior that increases considerably with the conformal weight of the corresponding bulk primary. These oscillations seem to limit us for example from determining the reflection and transmission coefficients that are associated to bulk fields of conformal weight $4$.

The OSFT approach is very versatile and can be implemented for different initial boundaries, defects or interfaces as long as their spectrum and the corresponding structure constants (or their subset) are known and contain finitely many fields at low levels. Note also that (compared to the RG analysis) the OSFT method allows for exploring conformal defects with higher $g$-function than the $g$-function of the initial defect. Such types of solutions were studied for example in \cite{Kudrna:2014rya,Kudrna:2019xnw} and require an analysis at higher levels.

\section*{Acknowledgements}

We would like to thank Davide Gaiotto, Katherine Latimer, Ingo Runkel, Martin Schnabl, Jakub Vo\v{s}mera and Jingxiang Wu for many interesting discussions and suggestions.  We are particularly grateful to Mat\v{e}j Kudrna for sharing his code that allowed higher-level calculations. We are thankful to Aiden Suter and Francisco A. Borges for collaboration in early stages of the project. The project started at a winter school of Perimeter Institute for Theoretical Physics organized by Lenka Bojdov\'{a} and Ma\"{i}te Dupuis. Research at Perimeter Institute is supported in part by the Government of Canada through the Department of Innovation, Science and Economic Development Canada and by the Province of Ontario through the Ministry of Economic Development, Job Creation and Trade.
The research of M.R. was supported by NSF grant 1521446, NSF grant 1820912, the Berkeley Center for Theoretical Physics and the Simons Foundation. J.M.R. thanks the Abdus Salam International Centre for Theoretical Physics, ICTP-SAIFR/IFT-UNESP and FAPESP grant 2016/01343-7 for partial financial support.

\appendix
\section{Defect structure constants}
\label{app:StructureConstants}

In order to determine the OSFT action (\ref{eq:OSFTaction}), we need defect structure constants. The structure constants for a subset of fields in the spectrum of the $D_{(1,2)}$ topological defect were first calculated in \cite{Makabe:2017uch}. For the convenience of reader, let us repeat their final results here with a few misprint corrections. The case of the Ising model ($m$=3) is slightly different compared to other minimal models, and it is presented at the end of this section.


We consider a subset of defect fields of the $D_{(1,2)}$ topological defect listed in table \ref{tab:spec12}. The defect structure constants $d_{ij}$ and $C_{ij}^k$ are defined in the following way
\begin{equation}
    \psi_i(x) \, \psi_j(y) = \frac{d_{ij}}{ (x-y)^{\Delta_i+\Delta_j}} + \frac{C_{ij}^{k}}{(x-y)^{\Delta_i+\Delta_j-\Delta_k}} \, \psi_k(y) + \dots \; ,
\end{equation}
where $\psi_i$ is a defect field of scaling dimension $\Delta_i=h_i+\bar{h}_i$. We will also need two bulk structure constants, $d_{\varphi\varphi}$ and $C^\varphi_{\varphi\varphi}$, of the diagonal bulk field $\varphi(z,\bar{z})$ with Kac label $(1,3)$ and conformal weights $(h,h)$ defined by
\begin{equation}
    \varphi(z,\bar{z}) \, \varphi(w,\bar{w}) = \frac{d_{\varphi\varphi}}{|z-w|^{4h}} + \frac{C_{\varphi\varphi}^{\varphi}}{|z-w|^{2h}} \, \varphi(w,\bar{w}) + \dots \; .
\end{equation}

\begin{table}[t]
\centering
\begin{tabular}{|c|c|c|}
\hline
\textbf{Notation} & \textbf{Primary defect field}                     & \textbf{Conformal weights}            \\ \hline
$\id$             & $\phi_{(1,1)(1,1)}$ & $\qty(0,0)$                 \\ \hline
$\phi$            & $\phi_{(1,3)(1,1)}$ & $\qty(h,\, 0)$         \\ \hline
$\bar{\phi}$      & $\phi_{(1,1)(1,3)}$ & $\qty(0,\, h)$         \\ \hline
$\varphi_L$       & $\phi_{(1,3)(1,3)}$ & $\qty(h,\, h)$ \\ \hline
$\varphi_R$       & $\phi_{(1,3)(1,3)}$ & $\qty(h,\, h)$ \\ \hline
\end{tabular}
\caption{Subset of defect spectrum of the $D_{(1,2)}$ defect. We adopt notation from \cite{Makabe:2017uch} and denote $h \equiv h_{1,3}=\frac{m-1}{m+1}$.}
\label{tab:spec12}
\end{table}
\FloatBarrier

We write the structure constants in terms of two-point functions $d_{\varphi\varphi}$ and $d_{\phi\phi}$. Normalization of all structure constants is fixed once we fix $d_{\varphi\varphi}$ and $d_{\phi\phi}$ \footnote{In this paper we use the normalization
\begin{align}
d_{\varphi\varphi} = \frac{\sin(3\pi t)}{\sin(\pi t)}, \qquad d_{\phi\phi} = - \frac{\Gamma(2-3t) \Gamma(2t)}{\Gamma(2-2t)\Gamma(t)} ,
\end{align}
where $t=\frac{m}{m+1}$. It corresponds to $\eta_{\Phi}=\eta_{\phi}=1$ in the convention of \cite{Runkel:2010ym}.}.
\subsubsection*{Two-point functions}
The non-zero two-point functions are equal
\begin{align}
    d_{\bar{\phi}\bar{\phi}} = d_{\phi\phi}, \quad d_{LL}=d_{RR}=d_{\varphi\varphi}, \quad d_{LR}=d_{RL}=\gamma d_{\varphi\varphi},
\end{align}
where $\gamma=2\cos(2\pi t)-1$ and $t=m/(m+1)$.

\subsubsection*{Three chiral fields ($\phi$ or $\bphi$)}
\begin{align}
 C^\phi_{\phi\phi} &= C^{\bphi}_{\bphi\bphi} =  \qty[d_{\phi\phi} \frac{\Gamma(2-3t)\Gamma(t)\Gamma(1-2t)^3}{\Gamma(2-4t)^2\Gamma(2t-1)\Gamma(1-t)^2}]^{\frac{1}{2}} \\
C^{\phi}_{\bphi\bphi} &= C^{\bphi}_{\phi\phi} = C^{\bphi}_{\phi\bphi} = C^{\bphi}_{\bphi\phi} = C^{\phi}_{\phi\bphi} = C^{\phi}_{\bphi\phi} = 0 \; .
\end{align}

\subsubsection*{Two chiral fields}

\begin{align}
C^L_{\phi\bphi} = \kappa \sqrt{\frac{d_{\phi\phi}^2}{d_{\varphi\varphi}(2 + \gamma\zeta + \gamma\zeta^{-1})}}, \qquad C^L_{\bphi\phi} = \kappa^{-1} \sqrt{\frac{d_{\phi\phi}^2}{d_{\varphi\varphi}(2 + \gamma\zeta + \gamma\zeta^{-1})}} ,
\end{align}
where the constants are
\begin{equation}
\zeta = e^{i \pi h}, \quad \kappa=\zeta^{1/2}.
\end{equation}
Then the rest can be written in terms of the two above.
\begin{align}
C^R_{\phi\bphi} &= C^L_{\bphi\phi}     &&      C^R_{\bphi\phi} = C^L_{\phi\bphi}  \\ 
C^{\bphi}_{R\phi} &= \frac{d_{\varphi\varphi}}{d_{\phi\phi}} \qty(C^L_{\bphi\phi} + \gamma \, C^L_{\phi\bphi})    &&  C^{\bphi}_{L\phi} = \frac{d_{\varphi\varphi}}{d_{\phi\phi}} \qty(\gamma\, C^L_{\bphi\phi} + C^L_{\phi\bphi})   \\ 
C^{\phi}_{R\bphi} &= \frac{d_{\varphi\varphi}}{d_{\phi\phi}} \qty(C^L_{\phi\bphi} + \gamma \, C^L_{\bphi\phi})    &&    C^{\phi}_{L\bphi} = \frac{d_{\varphi\varphi}}{d_{\phi\phi}} \qty(\gamma\, C^L_{\phi\bphi} + C^L_{\bphi\phi})   \\ 
C^{\bphi}_{\phi R} &= \zeta^{-2} \, \frac{d_{\varphi\varphi}}{d_{\phi\phi}}  \qty(C^L_{\bphi\phi} + \gamma \, C^L_{\phi\bphi})    &&    C^{\bphi}_{\phi L} = \zeta^2 \, \frac{d_{\varphi\varphi}}{d_{\phi\phi}}  \qty(\gamma\, C^L_{\bphi\phi} + C^L_{\phi\bphi})   \\ 
C^{\phi}_{\bphi R} &= \zeta^{2} \, \frac{d_{\varphi\varphi}}{d_{\phi\phi}}  \qty(C^L_{\phi\bphi} + \gamma \, C^L_{\bphi\phi})    &&    C^{\phi}_{\bphi L} = \zeta^{-2} \, \frac{d_{\varphi\varphi}}{d_{\phi\phi}}  \qty(\gamma\, C^L_{\phi\bphi} + C^L_{\bphi\phi})
\end{align} 

\vspace{-0.6cm}

\begin{align}
C^R_{\phi\phi} &= C^L_{\phi\phi} = C^R_{\bphi\bphi} = C^L_{\bphi\bphi} =0 \\
C^{\phi}_{R\phi} &= C^{\phi}_{L\phi} = C^{\bphi}_{R\bphi} = C^{\bphi}_{L\bphi} =0  \\
C^{\phi}_{\phi R} &= C^{\phi}_{\phi L} = C^{\bphi}_{\bphi R} = C^{\bphi}_{\bphi L} =0 \,.
\end{align}

\subsubsection*{One chiral field}
\begin{align}
 C^R_{\phi L}    = \frac{1+\gamma\zeta}{\zeta(1-\gamma^2)} C^{\phi}_{\phi\phi}, \qquad C^L_{\phi R}    = \frac{\gamma+\zeta}{1-\gamma^2} C^{\phi}_{\phi\phi} .
\end{align}
Then the rest can be written in terms of the two above.
\begin{align}
C^R_{\bphi L}   &= C^L_{\phi R} && C^L_{\bphi R}   = C^R_{\phi L} \\
C^R_{L\phi}     &= \zeta^{-1} \, C^R_{\phi L}    &&    C^R_{L\bphi}     = \zeta \, C^R_{\bphi L} \\
C^L_{R\phi}     &= \zeta \, C^L_{\phi R}    &&    C^L_{R\bphi}     = \zeta^{-1} \, C^L_{\bphi R} \\
C^L_{\phi L}    &= -\gamma \, C^R_{\phi L}       &&    C^R_{\phi R}     = -\gamma \, C^L_{\phi R} \\
C^L_{\bphi L}   &= -\gamma \, C^R_{\bphi L}      &&    C^R_{\bphi R}    = -\gamma \, C^L_{\bphi R} \\
C^L_{L\phi}     &= -\gamma\zeta^{-1} \, C^R_{\phi L} &&  C^L_{L\bphi}   = -\gamma\zeta \, C^R_{\bphi L}\\
C^R_{R\phi}     &= -\gamma\zeta \, C^L_{\phi R} &&  C^R_{R\bphi}   = -\gamma\zeta^{-1} \, C^L_{\bphi R}  \\
C^{\phi}_{LR}   &= \zeta (1-\gamma^2) \frac{d_{\varphi\varphi}}{d_{\phi\phi}}  \, C^L_{\phi R}  &&
C^{\bphi}_{LR}  = (1-\gamma^2) \frac{d_{\varphi\varphi}}{d_{\phi\phi}}  \, C^R_{\bphi L}\\
C^{\phi}_{RL}   &= (1-\gamma^2) \frac{d_{\varphi\varphi}}{d_{\phi\phi}}  \, C^L_{\phi R}  &&
C^{\bphi}_{RL}  = \zeta(1-\gamma^2) \frac{d_{\varphi\varphi}}{d_{\phi\phi}} \, C^R_{\bphi L} \\
\mathclap{ \hspace{8.5cm} C^{\phi}_{LL}=C^{\bphi}_{LL} =C^{\phi}_{RR} =C^{\bphi}_{RR} = 0 \,.}
\end{align}

\subsubsection*{No chiral fields}
The structure constant of the bulk field $\varphi(z,\bar{z})$ is
\begin{align}
C^\varphi_{\varphi\varphi} &= \qty[-d_{\varphi\varphi}(1-2t)^2 \frac{\Gamma(2-3t)}{\Gamma(3t-1)} \frac{\Gamma(4t-1)^2}{\Gamma(2-4t)^2} \frac{\Gamma(t)^3}{\Gamma(1-t)^3} \frac{\Gamma(1-2t)^4}{\Gamma(2t)^4}]^{\frac{1}{2}} .
\end{align}
Then the structure constants of defect fields $\varphi_L$ and $\varphi_R$ can be written using the above bulk structure constant:
\begin{align}
C^R_{RR}  &= C^L_{LL} = C_{\varphi\varphi}^{\varphi} \\
C^R_{LR} &= C^L_{LR} = C^R_{RL} = C^L_{RL} = \frac{\gamma}{1+\gamma} C^{\varphi}_{\varphi\varphi} \\
C^R_{LL} &= C^L_{RR} = 0 \,.
\end{align}

\subsection{Ising model} 

In case of the Ising model $\mathcal{M}_{4,3}$, there is only one copy of conformal family $\Phi\equiv\phi_{(1,3)(1,3)}$ in the $D_{(1,2)}$ spectrum. Therefore, the (closed) subset of conformal families from table \ref{tab:spec12} becomes $\qty{\id,\,\phi,\,\bphi,\,\Phi}$. The structure constant $C^\Phi_{\phi\bphi}=(C^\Phi_{\bphi\phi})^*$ was computed in \cite{Runkel:2010ym}. The only non-zero structure constants (up to the normalization of $d_{\varphi\varphi}$ and $d_{\phi\phi}$) are

\begin{alignat}{3}
d_{\bar{\Phi}\bar{\Phi}} &= d_{\varphi\varphi} && \qquad\qquad d_{\bar{\phi}\bar{\phi}} &&= d_{\phi\phi} \\
C^{\Phi}_{\phi\bphi} &= \frac{d_{\phi\phi}}{\sqrt{d_{\varphi\varphi}}} \; i && \qquad\qquad C^{\Phi}_{\bphi\phi} &&= - \, \frac{d_{\phi\phi}}{\sqrt{d_{\varphi\varphi}}} \; i \\
C^{\phi}_{\bphi\Phi} &= C^{\bphi}_{\Phi\phi} = \sqrt{d_{\varphi\varphi}} \; i && \qquad\qquad C^{\phi}_{\Phi\bphi} &&= C^{\bphi}_{\phi\Phi} = - \, \sqrt{d_{\varphi\varphi}} \; i \, .
\end{alignat}

\section{$g$-functions of OSFT solutions}
\label{app:Tables}

In this appendix we show the $g$-functions of the numerical OSFT solutions corresponding to defects $D_{(2,1)}$, $D_{(1,1)}$, $F$ and $C$ in tables \ref{tab:gD21}, \ref{tab:gD11}, \ref{tab:gF} and \ref{tab:gC}, respectively. The extrapolations to the infinite level are achieved by fitting a quadratic function in $1/L$ using results at integral levels and sending $L\rightarrow\infty$.

\begin{table}[h!]
\begin{tabular}{|c|c|c|c|c|c|c|c|c|}
\hline
level    & $m=4$                & $m=5$   & $m=6$   & $m=7$   & $m=8$   & $m=9$   & $m=10$  & $m=100$ \\ \hline
1.0      & 1.49951              & 1.66355 & 1.75839 & 1.81821 & 1.85835 & 1.88658 & 1.90717 & 1.99901 \\
1.5      & 1.49951              & 1.66355 & 1.75839 & 1.81821 & 1.85835 & 1.88658 & 1.90717 & 1.99901 \\
2.0      & 1.45012              & 1.63900 & 1.74474 & 1.80997 & 1.85306 & 1.88301 & 1.90467 & 1.99901 \\
2.5      & 1.45012              & 1.63900 & 1.74474 & 1.80997 & 1.85306 & 1.88301 & 1.90467 & 1.99901 \\
3.0      & 1.43839              & 1.63130 & 1.73980 & 1.80673 & 1.85088 & 1.88149 & 1.90359 & 1.99901 \\
3.5      & 1.43839              & 1.63130 & 1.73980 & 1.80673 & 1.85088 & 1.88149 & 1.90359 & 1.99901 \\
4.0      & 1.43022              &         &         & 1.80534 & 1.84999 & 1.88090 & 1.90318 & 1.99901 \\
4.5      & 1.43022              &         &         &         &         &         &         &         \\
5.0      & 1.42732              &         &         &         &         &         &         &         \\
5.5      & 1.42732              &         &         &         &         &         &         &         \\
6.0      & 1.42428              &         &         &         &         &         &         &         \\ \hline
$\infty$ & 1.41347              & 1.61665 & 1.72933 & 1.80027 & 1.84655 & 1.87851 & 1.90148 & 1.99901 \\ \hline
$D_{(2,1)}$    & 1.41421              & 1.61803 & 1.73205 & 1.80194 & 1.84776 & 1.87939 & 1.90211 & 1.99901 \\
\hline
\end{tabular}
\caption{The $g$-functions of the OSFT solution corresponding to defect $D_{(2,1)}$.}
\label{tab:gD21}
\end{table}

\begin{table}[h!]
\begin{tabular}{|c|c|c|c|c|c|c|c|c|}
\hline
level    & $m=4$                & $m=5$   & $m=6$   & $m=7$   & $m=8$   & $m=9$   & $m=10$  & $m=100$ \\ \hline
1.0      & 1.28760              & 1.36574 & 1.42267 & 1.46596 & 1.49997 & 1.52741 & 1.54999 & 1.75570 \\
1.5      & 1.28760              & 1.36574 & 1.42267 & 1.46596 & 1.49997 & 1.52741 & 1.54999 & 1.75570 \\
2.0      & 1.12303              & 1.17445 & 1.21531 & 1.24813 & 1.27490 & 1.29710 & 1.31576 & 1.50048 \\
2.5      & 1.12303              & 1.17445 & 1.21531 & 1.24813 & 1.27490 & 1.29710 & 1.31576 & 1.50048 \\
3.0      & 1.08493              & 1.11807 & 1.14504 & 1.16711 & 1.18537 & 1.20067 & 1.21365 & 1.34823 \\
3.5      & 1.08493              & 1.11807 & 1.14504 & 1.16711 & 1.18537 & 1.20067 & 1.21365 & 1.34823 \\
4.0      & 1.05597              &         &         & 1.12428 & 1.14029 & 1.15385 & 1.16543 & 1.28770 \\
4.5      & 1.05597              &         &         &         &         &         &         &         \\
5.0      & 1.04732              &         &         &         &         &         &         &         \\
5.5      & 1.04732              &         &         &         &         &         &         &         \\
6.0      & 1.03625              &         &         &         &         &         &         &         \\ \hline
$\infty$ & 0.999882             & 1.01636 & 1.00279 & 0.990664 & 0.990343 & 0.990563 & 0.991071 & 1.01083 \\ \hline
$D_{(1,1)}$    & 1                    & 1 & 1 & 1 & 1 & 1 & 1 & 1 \\ \hline
\end{tabular}
\caption{The $g$-functions of the OSFT solution corresponding to defect $D_{(1,1)}$.}
\label{tab:gD11}
\end{table}

\begin{table}[h!]
\begin{tabular}{|c|c|c|c|c|c|c|c|c|}
\hline
level    & $m=4$                & $m=5$   & $m=6$   & $m=7$   & $m=8$   & $m=9$   & $m=10$  & $m=100$ \\ \hline
1.0      & 1.32839              & 1.43289 & 1.50736 & 1.56283 & 1.60561 & 1.63952 & 1.66701 & 1.89164 \\
1.5      & 1.31078              & 1.40791 & 1.47944 & 1.53390 & 1.60561 & 1.63952 & 1.66701 & 1.89164 \\
2.0      & 1.19741              & 1.28010 & 1.34519 & 1.39688 & 1.43855 & 1.47268 & 1.50106 & 1.76067 \\
2.5      & 1.15521              & 1.23482 & 1.29872 & 1.35011 & 1.43855 & 1.47268 & 1.50106 & 1.76067 \\
3.0      & 1.13233              & 1.20432 & 1.26219 & 1.30886 & 1.34694 & 1.37843 & 1.40482 & 1.65703 \\
3.5      & 1.12920              & 1.20033 & 1.25717 & 1.30298 & 1.34694 & 1.37843 & 1.40482 & 1.65703 \\
4.0      & 1.10711              &         &         & 1.27293 & 1.30929 & 1.33950 & 1.36494 & 1.61263 \\
4.5      & 1.10104              &         &         &         &         &         &         &         \\
5.0      & 1.09547              &         &         &         &         &         &         &         \\
5.5      & 1.09458              &         &         &         &         &         &         &         \\
6.0      & 1.08596              &         &         &         &         &         &         &         \\ \hline
$\infty$ & 1.02184              & 1.01545 & 1.05275 & 1.11797 & 1.14567 & 1.16915 & 1.18926 & 1.40470 \\ \hline
$F$    & 1.05146              & 1.09819 & 1.13617 & 1.16677 & 1.19166 & 1.21218 & 1.22934 & 1.39327 \\ \hline
\end{tabular}
\caption{The $g$-functions of the OSFT solution corresponding to defect $F$.}
\label{tab:gF}
\end{table}

\begin{table}[h!]
\begin{tabular}{|c|c|c|c|c|c|c|c|c|}
\hline
level    & $m=4$                & $m=5$   & $m=6$   & $m=7$   & $m=8$   & $m=9$   & $m=10$  & $m=100$ \\ \hline
1.0      &  1.47897          & 1.64357 & 1.74202 & 1.80523 & 1.84807 & 1.87836 & 1.90052 & 1.99900 \\
1.5      & 1.40544 & 1.61430 & 1.72942 & 1.79861 & 1.84807 & 1.87836 & 1.90052 & 1.99900 \\
2.0      & 1.29648              & 1.55802 & 1.70454 & 1.78509 & 1.83575 & 1.87024 & 1.89489 & 1.99900 \\
2.5      & 1.29297              & 1.55534 & 1.70153 & 1.78279 & 1.83575 & 1.87024 & 1.89489 & 1.99900 \\
3.0      & 1.25647              & 1.53317 & 1.69157 & 1.77719 & 1.83053 & 1.86663 & 1.89232 & 1.99899 \\
3.5      & 1.23089              & 1.51431 & 1.68726 & 1.77578 & 1.83053 & 1.86663 & 1.89232 & 1.99899 \\
4.0      & 1.20140              &         &         & 1.77260 & 1.82799 & 1.86503 & 1.89123 & 1.99899 \\
4.5      & 1.20111              &         &         &         &         &         &         &         \\
5.0      & 1.18937              &         &         &         &         &         &         &         \\
5.5      & 1.17871              &         &         &         &         &         &         &         \\
6.0      & 1.16533              &         &         &         &         &         &         &         \\ \hline
$\infty$ & 1.09852              & 1.48896 & 1.66489 & 1.75860 & 1.81910 & 1.85896 & 1.88692 & 1.99899\\ \hline
\end{tabular}
\caption{The $g$-functions of the OSFT solution corresponding to defect $C$. The solution at level $L=1.5$ for $m=4$ was found by solving the equations for saddle points at that level. A more precise extrapolation of the $g$-function of defect $C$ in the tricritical Ising model, using even, odd and half-integer levels separately, is given by (\ref{eq:gCm4}).}
\label{tab:gC}
\end{table}


\bibliographystyle{JHEP}
\bibliography{References}

\end{document}